\documentclass[aps,prl,floatfix,showpacs,twocolumn,preprintnumbers,tightenlines,superscriptaddress]{revtex4-1}

\usepackage{papersstyle}

\newcommand{\erf}{\text{erf}}

\newcommand{\Hrabi}{\ensuremath{H_{\rm R}}}
\newcommand{\Hrabiint}{\ensuremath{H_{\rm R, int}}}
\newcommand{\Hjc}{\ensuremath{H_{\rm JC}}}
\newcommand{\Hajc}{\ensuremath{H_{\rm AJC}}}

\newcommand{\Trabi}{T^{\rm R}}
\newcommand{\Tjc}{T^{\rm JC}}

\newcommand{\Urabi}{\ensuremath{U_{\rm R}}}
\newcommand{\UTrRabi}{\ensuremath{\Urabi^{\rm Tr}}}
\newcommand{\Ujc}{\ensuremath{U_{\rm JC}}}
\newcommand{\Uajc}{\ensuremath{U_{\rm AJC}}}
\newcommand{\Rgen}[2]{R_{\rm #1}\!\br{#2}}
\newcommand{\dphi}{\Delta\phi}

\newcommand{\ketq}[1]{|#1\rangle_{\rm q}}
\newcommand{\ketr}[1]{|#1\rangle_{\rm r}}
\newcommand{\ketqr}[1]{|#1\rangle_{\rm q, r}}

\newcommand{\QR}{\ensuremath{Q_{\rm R}}}
\newcommand{\QW}{\ensuremath{Q_{\rm W}}}
\newcommand{\RR}{\ensuremath{R_{\rm R}}}

\newcommand{\rr}{}
\newcommand{\qw}{_{\rm W}}

\newcommand{\omq}{\omega_{\rm q}}

\newcommand{\omr}{\omega_{\rm r}}
\newcommand{\omrf}{\omega_{\rm RF}}

\newcommand{\Delq}{\Delta_{\rm q}}

\newcommand{\Delr}{\Delta_{\rm r}}
\newcommand{\Delqr}{\Delta_\text{q--r}}

\newcommand{\omqR}{\omega_{\rm q}^{\rm R}}
\newcommand{\omrR}{\omega_{\rm r}^{\rm R}}
\newcommand{\gR}{g^{\rm R}}
\newcommand{\gomratio}{r}

\newcommand{\omqzeroone}{\omega_{\rm q}^\text{0--1}}
\newcommand{\omqonetwo}{\omega_{\rm q}^\text{1--2}}

\newcommand{\be}{\begin{align}}
\newcommand{\ee}{\end{align}}

\newcommand{\undertilde}[1]{\rlap{\smash{\lower 2ex \hbox{$\tilde{\hphantom{#1}}$}}}#1}
\newcommand{\lundertilde}[1]{\rlap{\smash{\lower 2.2ex \hbox{$\tilde{\hphantom{#1}}$}}}#1}
\newcommand{\br}[1]{\left(#1\right)}		
\newcommand{\sqbr}[1]{\left[#1\right]}		

\newcommand{\bra}[1]{\langle#1|}		
\newcommand{\ket}[1]{|#1\rangle}		

\newcommand{\tr}{\mathrm{Tr}}

\newcommand{\sigz}{\sigma_{\rm z}}
\newcommand{\sigy}{\sigma_{\rm y}}
\newcommand{\sigx}{\sigma_{\rm x}}
\newcommand{\sigp}{\sigma^{+}}
\newcommand{\sigm}{\sigma^{-}}
\newcommand{\sigpm}{\sigma^{\pm}}

\newcommand{\ad}{a^\dagger}
\newcommand{\ada}{a^\dagger a}

\newcommand{\Techo}{T_{2,\mathrm{echo}}}
\newcommand{\Tone}{T_{1}}
\newcommand{\Toneres}{T_{\rm 1, r}}

\newcommand{\Ttwostar}{T_{2}^{\ast}}

\newcommand{\K}{\ensuremath{\mathrm{K}}}
\newcommand{\mK}{\ensuremath{\mathrm{mK}}}

\newcommand{\kHz}{\ensuremath{\mathrm{kHz}}}
\newcommand{\MHz}{\ensuremath{\mathrm{MHz}}}
\newcommand{\GHz}{\ensuremath{\mathrm{GHz}}}

\newcommand{\us}{\ensuremath{\mu\mathrm{s}}}
\newcommand{\ns}{\ensuremath{\mathrm{ns}}}

\newcommand{\dB}{\ensuremath{\mathrm{dB}}}

\newcommand{\EJmax}{\ensuremath{E_{\rm J, max}}}
\newcommand{\EJmin}{\ensuremath{E_{\rm J, min}}}

\newcommand{\Rmnum}[1]{\expandafter\@slowromancap\romannumeral #1@}

\begin{document}

\title{Experimentally simulating the dynamics of quantum light and matter at ultrastrong coupling}

\author{N.~K.~Langford}
\author{R.~Sagastizabal}
\author{M.~Kounalakis}
\author{C.~Dickel}
\author{A.~Bruno}
\author{F.~Luthi}
\affiliation{QuTech, Delft University of Technology, Delft, The Netherlands}
\affiliation{Kavli Institute of Nanoscience, Delft University of Technology, Lorentzweg 1, 2628 CJ Delft, The Netherlands}
\author{D.~J.~Thoen}
\author{A.~Endo}
\affiliation{Department of Microelectronics, Faculty of Electrical Engineering, Mathematics and Computer Science, Delft University of Technology, Mekelweg 4, 2628 CD Delft, The Netherlands}
\affiliation{Kavli Institute of Nanoscience, Delft University of Technology, Lorentzweg 1, 2628 CJ Delft, The Netherlands}
\author{L.~DiCarlo}
\affiliation{QuTech, Delft University of Technology, Delft, The Netherlands}
\affiliation{Kavli Institute of Nanoscience, Delft University of Technology, Lorentzweg 1, 2628 CJ Delft, The Netherlands}

\begin{abstract}
The quantum Rabi model describing the fundamental interaction between light and matter is a cornerstone of quantum physics.
It predicts exotic phenomena like quantum phase transitions and ground-state entanglement in the ultrastrong-coupling (USC) regime, where coupling strengths are comparable to subsystem energies.
Despite progress in many experimental platforms, the few experiments reaching USC have been limited to spectroscopy: demonstrating USC dynamics remains an outstanding challenge.
Here, we employ a circuit QED chip with moderate coupling between a resonator and transmon qubit to realise accurate digital quantum simulation of USC dynamics.
We advance the state of the art in solid-state digital quantum simulation by using up to 90 second-order Trotter steps and probing both subsystems in a combined Hilbert space dimension $\sim80$, demonstrating the Schr\"odinger-cat like entanglement and build-up of large photon numbers characteristic of deep USC.
This work opens the door to exploring extreme USC regimes, quantum phase transitions and many-body effects in the Dicke model.
\end{abstract}

\maketitle

Digital quantum simulations~\cite{Lloyd96} promise a quantum advantage without a universal, fault-tolerant quantum computer, with applications in fields such as quantum chemistry~\cite{Kassal08, OMalley16}
and condensed-matter physics~\cite{Abrams97, Lanyon11, Salathe15, Barends15}.
In a digital quantum simulator, sequences of discrete interaction components synthesise the evolution of an artificial Hamiltonian, allowing access to more exotic dynamics than the simulator can realise naturally.
Systems involving ultrastrong light-matter interactions raise significant challenges for both theoretical analysis~\cite{Rabi36, Ciuti05, Braak11, Beaudoin11, Nataf10, Viehmann11} and experimental study~\cite{Lolli15}, making them ripe candidates for exploration via quantum simulation.

Ultrastrong light-matter coupling~\cite{Ciuti05} has been achieved in a range of physical systems, including circuit quantum electrodynamics (QED)~\cite{Niemczyk10, Forn-Diaz16, Yoshihara16}, semiconductor quantum-well systems~\cite{Gunter09}, terahertz electron cyclotron transitions~\cite{Scalari12, Maissen14, Zhang16} and photochromic molecules~\cite{Schwartz11}.
While some experiments have now demonstrated spectroscopic signatures of deep ultrastrong coupling (dUSC)~\cite{Maissen14, Forn-Diaz16, Yoshihara16}, where the coupling-to-frequency ratio $g/\omega \gtrsim 1$, a dynamical signature has only been measured at $g/\omega \sim 0.09$~\cite{Zhang16}.
Theory suggests that simulations of the quantum Rabi model (QRM) could explore widely varied coupling regimes in architectures like circuit QED~\cite{Mezzacapo14, Ballester12, Lamata16}, cold atoms~\cite{Felicetti16} and trapped ions~\cite{Pedernales15}.
A classical analog simulation of evolution in a restricted subspace of the QRM has been performed in photonic waveguide systems~\cite{Crespi12, Casanova10}.

The standard QRM~\cite{Rabi36} describes the dynamics of a single two-level atom (energy $\hbar\omqR$) coupled to a single quantum harmonic field mode (energy $\hbar\omrR$) by field-dipole interaction:
\begin{align}
\nonumber
\frac{\Hrabi}{\hbar} = -\frac{\omqR}{2} \sigz + \omrR \ada + \gR \br{a+\ad}\br{\sigp + \sigm},
\end{align}
where $a$ ($\ad$) and $\sigm$ ($\sigp$) are annihilation (creation) operators for field mode and atom, respectively, $\sigma_{\rm j}$ are the Pauli spin operators, and $\gR$ is the coupling strength.
Under small coupling ($\gR \ll \omqR, \omrR$), this reduces to the Jaynes-Cummings (JC) model via the rotating-wave approximation:
\begin{align}
\nonumber
\frac{\Hjc}{\hbar} = -\frac{\omq}{2} \sigz + \omr \ada + g \br{a \sigp + \ad\sigm},
\end{align}
which contains only the excitation-number-conserving interaction terms, $a\sigp$ and $\ad\sigm$,
and has an exact solution.
In the USC regime ($\gR \sim \omqR, \omrR$), however, the excitation-nonconserving terms $a\sigm$ and $\ad\sigp$ cannot be neglected and only total parity [$\sigz\sum_n ({-}1)^n \ket{n}\bra{n}$] is conserved~\cite{Casanova10}.
Without the strong symmetry of number conservation, the combination of an infinite-dimensional oscillator with an explicitly quantum (two-level) atomic system makes the full QRM difficult to solve~\cite{Braak11}.
Demonstrating the ground-state entanglement and large ground-state photon numbers that can arise in the QRM is an open challenge in USC research.

\begin{figure*}[t]
  \begin{center}
  \includegraphics[width=\linewidth]{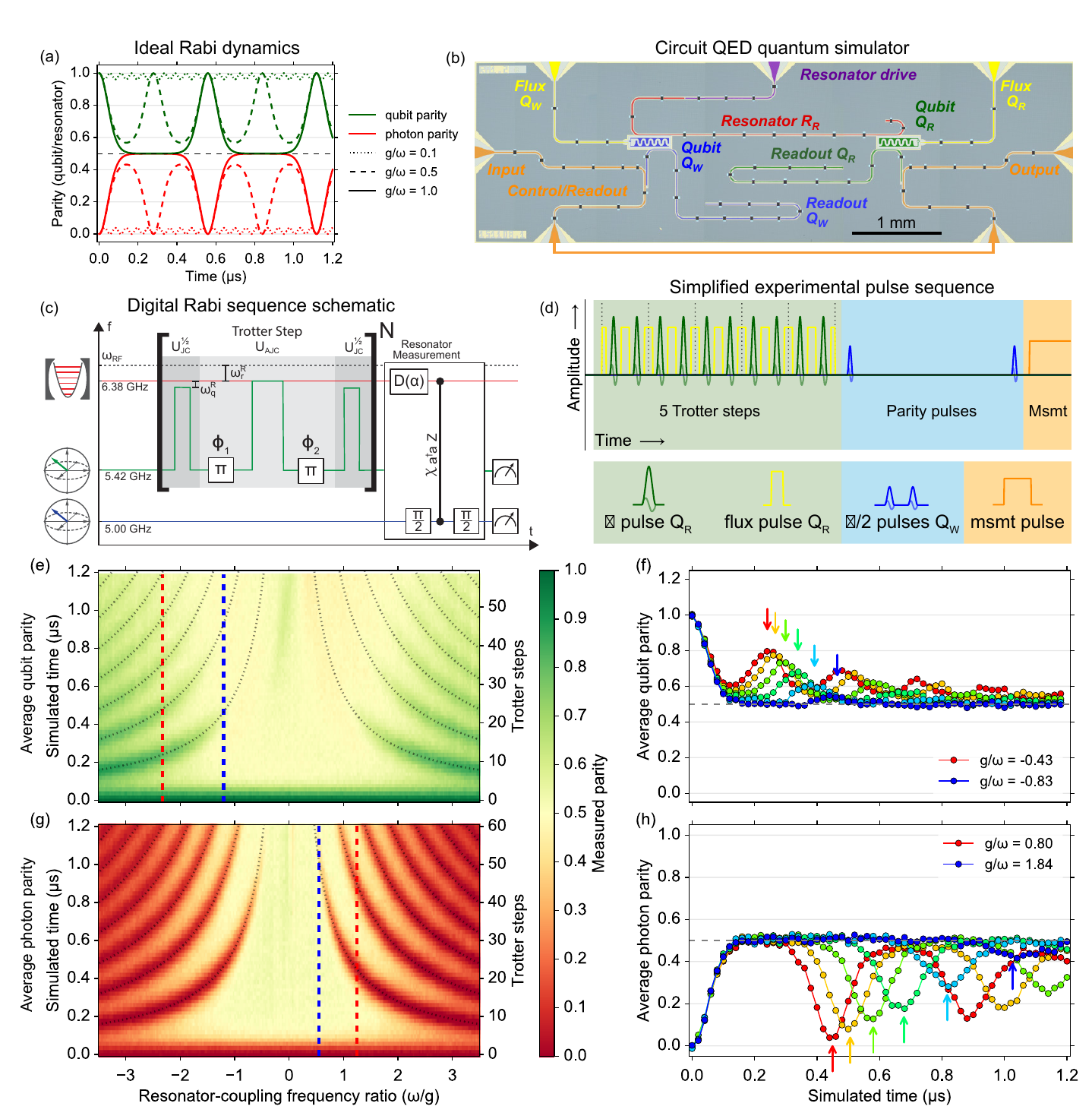}  
  \end{center}
  \caption{
  {\bf Digital quantum simulation of quantum Rabi model parity dynamics in the degenerate-qubit case.}
  (a) Parity dynamics of the ideal quantum Rabi model for qubit (green) and resonator (red) in coupling regimes: $\gomratio=\gR/\omqR=0.1$ (dotted), $0.5$ (dashed) and $1.0$ (solid).
  (b) Two-transmon, three-resonator circuit QED chip (detailed description in supplement~\cite{SOM}).
  (c) Sequence schematic for second-order Trotterisation.  The rotating frame defining the simulated resonator frequency ($\omr$) is controlled via the $\QR$ bit-flip pulse phases.
  (d) Example simplified experimental pulse sequence for 5 Trotter steps followed by a photon parity measurement.
  (e--h) Measured qubit and photon parity dynamics for up to 60 Trotter steps, with the extreme dUSC regime in the centre decreasing to weaker USC near the edges.
  The data show clear Gaussian-shaped collapses for all $\gomratio$, along with the characteristic plateaus of the ultrastrong coupling regime.
Qubit revivals are observed up to $\gomratio\sim0.8$, while photon parity shows clear revivals up to $\gomratio\sim1.8$.  Slices are plotted in (f, h) for evenly spaced $\omrR/\gR$ between the red and blue dashed lines in (e, g), respectively.
For $\gomratio\gtrsim1.5$, some deviation from the expected revival time in the photon parity results from a small residual Kerr-type nonlinearity in the resonator~\cite{SOM} and is correlated with significant photon populations.
  Arrows in (f) and (h) show expected revival times for each slice.
  }
  \label{fig:QubitVsParity}
\end{figure*}

Ultrastrong-coupling dynamics can also produce nontrivial quantum states and significant build-up of photon numbers~\cite{Casanova10}.
Many characteristic dynamical features of USC can already be seen in the degenerate-qubit limit $\omqR=0$.
Here, the interaction-picture Hamiltonian
\begin{align}
\nonumber
\Hrabiint = \sigx \br{\hbar \gR e^{-i\omrR t} a+\hbar \gR e^{i\omrR t}\ad}
\end{align}
is a coherent drive on the oscillator mode, with an amplitude $\pm \gR e^{i\omrR t}$ conditioned on the $\sigx$ basis state of the atom.
The conditional coupling $\pm\gR$ coherently displaces the field, but in a continuously rotating direction given by $e^{i\omrR t}$, creating two diametrically opposite circular trajectories in phase space [see Fig.~\ref{fig:ResoDynamics}(a)].
Relating the diameter and circumference of these trajectories, $\pi\alpha_{\rm max} = \dot{\alpha} \Trabi$, with field displacement rate $\dot{\alpha}=\gR$ and period $\Trabi=2\pi/\omrR$, gives a maximum amplitude $\alpha_{\rm max} = 2r$ set by the relative coupling ratio $\gomratio \equiv \gR/\omrR$.
Figure~\ref{fig:QubitVsParity}(a) illustrates the atomic and photonic parity dynamics for characteristic USC regimes, starting in an eigenstate of the uncoupled system, $\ketq{1}\otimes\ketr{0}$.
Because this is a superposition of $\sigx$ eigenstates, evolution gives rise to an atom-field entangled state (Bell-cat state)~\cite{Vlastakis15}, $\ketqr{+,{+}\alpha}-\ketqr{-,{-}\alpha}$.
For $\gomratio\ll1$, the two trajectories remain virtually indistinguishable, giving evolution closely approximating simple JC dynamics with an atom-field detuning equal to $\omrR$~\cite{SOM}.
As $\gomratio$ increases, the curves start distorting from the sinusoidal JC exchange oscillations (USC regime), until dUSC is reached, where the parities exhibit a characteristic Gaussian-shaped ``collapse'', followed by flat plateaus and periodic revivals at multiples of $\Trabi$.
The cross-over between these dynamical regimes is related to the maximum distinguishability of the two coherent states of the field.
When the paths separate completely, the qubit appears to be in a mixed state, with parity 0.5.

In our circuit QED simulator, the Rabi atom and field mode are simulated by a transmon qubit (\QR)~\cite{Koch07} and a coplanar waveguide resonator (\RR) with energies $\hbar\omq$ and $\hbar\omr$, respectively [device shown in Fig.~\ref{fig:QubitVsParity}(b)].
Because the transmon is only weakly anharmonic ($\omqzeroone{-}\omqonetwo \ll \omqzeroone$), directly increasing the qubit-resonator coupling $g$ leads to a breakdown in its qubit behaviour at small $\gomratio$, and full circuit quantization shows that dUSC cannot be reached for any circuit parameters~\cite{Jaako16}.
Instead, building on the proposal in Ref.~\onlinecite{Mezzacapo14}, we perform a digital simulation of the QRM for arbitrarily large $\gomratio$ using a coupling in the manifestly non-USC regime ($\gomratio < 10^{-3}$).
The full Rabi Hamiltonian can be decomposed into two JC-like interactions~\cite{Mezzacapo14}:
\begin{align}
\nonumber
\Hrabi(\gR, \omrR, \omqR)=\Hjc(g, \Delr, \Delq^{\rm JC}) + \Hajc(g, \omr, \Delq^{\rm AJC}),
\end{align}
where $\Hajc=\sigx \Hjc \sigx$ contains only counter-rotating interaction terms, and the effective Rabi parameters $\gR=g$, $\omrR=2\Delr$ and $\omqR=\Delq\equiv\Delq^{\rm JC}-\Delq^{\rm AJC}$ are not related to the natural circuit frequencies, but defined relative to a nearby rotating frame ($\Delta=\omega-\omrf$), and can be arbitrarily small.
Using the standard method of Trotterization~\cite{Lloyd96}, Rabi dynamics can therefore be simulated deep into the USC regime by decreasing $\Delr$ and $\Delq$.
Figure~\ref{fig:QubitVsParity}(c) illustrates the second-order Trotter step used here (see Methods and~\cite{SOM}).
An asymmetric transmon with two flux-insensitive ``sweet'' spots~\cite{Koch07, SOM} is driven and measured at its lower sweet spot (5.452~\GHz) far below the resonator (6.381~\GHz), and short JC interaction gates are applied by fast frequency-tuning flux pulses~\cite{DiCarlo09}.
Experimentally, a rotating frame is usually defined by the frequency of a drive tone.
Here, the choice of rotating frame specifies the required rotation axis of the $\pi$ pulses which create the AJC interaction.
By appropriately updating the pulse phases, which are controlled with high precision, we can therefore arbitrarily select the rotating frame detuning from the resonator, even though these pulses are applied far from both resonator and rotating frame (see Methods).

Numerical modelling of the digital Rabi protocol highlighted several challenges for device design and fabrication~\cite{SOM}.
Most significantly, due to practical flux-pulsing bandwidths which limit the shortest achievable Trotter step, it is challenging to digitise fast compared with the dynamics.
Reaching acceptably low Trotter error in interesting regimes of $\gomratio$ therefore required small qubit-resonator coupling (here, $g/2\pi=1.95$~\MHz).
This also placed constraints on other device parameters, including coherence (for long simulation times), flux-tuning precision and qubit-resonator frequency targetting (due to a very narrow resonance).
An extra qubit \QW\ was strongly and dispersively coupled to \RR\ to probe the intraresonator quantum state via its photon-dependent frequency shift ($-1.26$~\MHz\ per photon) using pulse sequences based on Ramsey interferometry.
We used \QW\ to implement a range of photon measurements: average photon number with a controllable dynamic range, average photon parity and, combined with coherent field displacements through an external input coupler, direct Wigner tomography of the resonator~\cite{SOM, Bertet02, Riste13, Vlastakis13}. 
Qubits were driven and measured through dedicated read-out resonators.

\begin{figure*}[t]
  \centering
  \includegraphics[width=\linewidth]{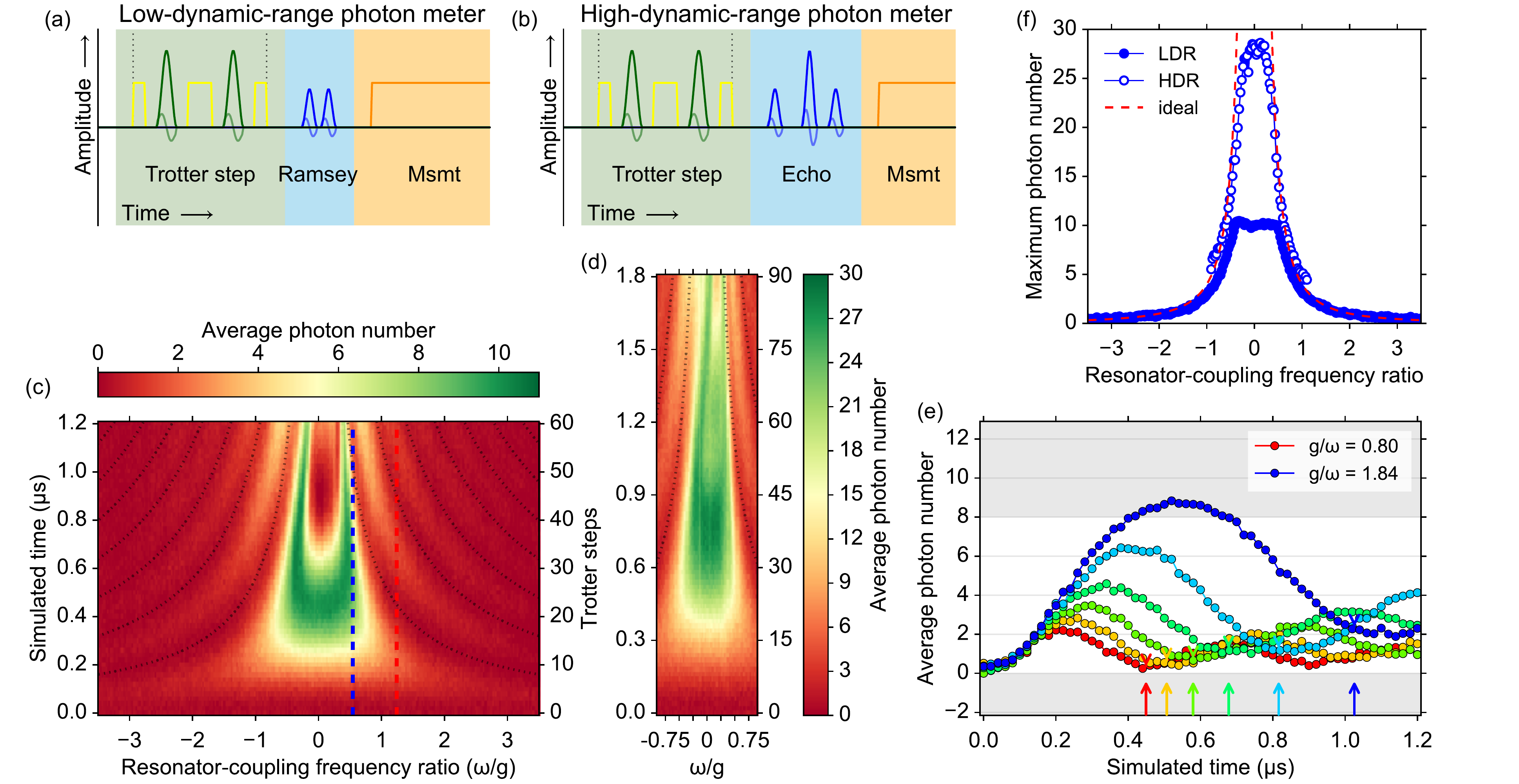}
  \caption{
  {\bf Photon number dynamics of the quantum Rabi model in the degenerate-qubit case.}
  (a, b) Average photon number is probed by applying Ramsey and echo-like pulses to $\QW$.  The effective Ramsey pulse separation $\tau$ determines the photon dynamic range.  Because of finite pulse widths, reaching the small $\tau$ needed for high dynamic ranges (b) requires an unbalanced ``echo''-like sequence.  
  (c, e) Measured photon number dynamics up to 60 Trotter steps using a low-dynamic-range (LDR) photon number meter ($\tau\sim18.7$ \ns) with a linear range of $\sim$ 0--8 photons [indicated by grey regions in (e)].
  Large photon populations in the resonator highlight the non-conservation of excitation number in the quantum Rabi model.  The resonator displays clear oscillations up to $\gomratio>1.8$ in good agreement with the expected qubit revival times (dotted curves).  The red feature in the middle reflects the upper limit on the photon meter's dynamic range set by $\QW$ ``population wrapping'' at high photon numbers.
  (d) Measured photon dynamics up to 90 Trotter steps using a high-dynamic-range (HDR) photon meter with $\tau\sim6.5$ \ns\ and a linear range of $\sim$ 0--20 photons, allowing observation of photon oscillations beyond 1.5 \us\ of simulated time (more than 75 Trotter steps).  This data shows the effect of a residual Kerr nonlinearity at high values of $\gomratio$.
  (e) Line slices are plotted for evenly spaced resonator-coupling frequency ratios between the red and blue dashed lines shown in (c).  Grey regions delineate the linear range of the photon meter.
  (f) Maximum measured average photon number for each value of $\gomratio$ for both LDR and HDR photon meters.
  }
  \label{fig:PhotonNumberChevrons}
\end{figure*}

We first experimentally simulate the QRM for the degenerate-qubit case over a wide range of $\gomratio$ covering the full USC regime, from $\gomratio\sim0.3$ to $\gomratio\rightarrow\infty$ (Fig.~\ref{fig:QubitVsParity}).
We use 60 Trotter steps to simulate 1.2~\us\ of dynamics ($gt=4.68\pi$) and measure either qubit or photon parity after each step.
(Simulations start in the state $\ketq{1}\otimes\ketr{0}$ for all results in the main text.)
A simplified pulse sequence is illustrated in Fig.~\ref{fig:QubitVsParity}(d).
The qubit and photon parity dynamics [Fig.~\ref{fig:QubitVsParity}(e, g)] show very similar qualitative behaviour, consistent with parity conservation.
The revival periods $\Trabi$ are in excellent agreement with the predictions of USC Rabi dynamics, and strikingly different from those predicted for a pure Jaynes-Cummings interaction with the equivalent qubit-resonator detuning $\br{\Tjc=2\pi/\sqrt{4g^2+\Delqr^2}}$~\cite{SOM}.
The experimental simulations also show the Gaussian-shaped parity collapse set by the simulated $\gR$, which is a key signature of dUSC dynamics.
From fits to the initial points of the qubit data, we calculate an average $\gR\approx2\pi\times1.79$~\MHz, slightly lower than the expected $\gR=g\approx2\pi\times1.95$~\MHz\ determined from independent spectroscopy and vacuum Rabi oscillations.
This is consistent with a small residual flux-pulse distortion and provides the best estimate for the simulated $\gR$ achieved in these experiments.

From the observation of parity revivals, combined with the simulated $\gR$, we can estimate the range of $\gomratio$ reached in these simulations.
For $\gR/2\pi=1.79$~\MHz\ and $\gomratio=1$ (archetypal dUSC), the expected revival time is 0.56~\us.
Line slices for the qubit parity dynamics [Fig.~\ref{fig:QubitVsParity}(f)] show revivals beyond 0.4~\us\ ($\gomratio\sim0.7$).
Photon parity revivals, however, persist beyond 1.0~\us\ ($\gomratio\sim1.8$).
This difference again results from photon decay, as shown by excellent agreement with numerical modelling which includes cavity decay but no other decoherence~\cite{SOM}.
Photon decay becomes increasingly critical in the USC regime, because even a single decay destroys the qubit-resonator entanglement, and losing a photon becomes increasingly likely for larger photon numbers.
The qubit parity revivals rely on entanglement being maintained.
This is supported by measurements of reduced qubit entropy, which show that the qubit state collapses to the mixed state, before displaying a revival in purity~\cite{SOM}.
The resonator parity dynamics, however, are more robust to decay and provide a more direct measure of the dUSC dynamics.
Photon parity collapses and revivals prove the field undergoes large-amplitude excursions through phase space even during a single cycle of the resonator period.
The difference between qubit and photon parity dynamics is a signature of break-down in parity conservation, caused by resonator decay.

We next directly explore the build-up of large photon populations (Fig.~\ref{fig:PhotonNumberChevrons}), another feature of USC dynamics that contrasts strikingly with the excitation-conserving dynamics expected under weak coupling.
Using a Ramsey pulse sequence with small separation $\tau$, the excitation probability in $\QW$ becomes a measure of average photon number in the resonator~\cite{SOM}.
The dynamic range and sensitivity of this photon meter are controlled via $\tau$ [Figs~\ref{fig:PhotonNumberChevrons}(a, b)].
Measured with a linear range of $\sim$ 0--8 photons [Fig.~\ref{fig:PhotonNumberChevrons}(c)], the resonator displays the complementary build-up of photons which causes the collapse of qubit and photon parity, clearly demonstrating the violation of number conservation expected for the QRM.
As with photon parity, clear oscillations can be seen out to $\gomratio\sim1.8$ [Fig.~\ref{fig:PhotonNumberChevrons}(e)].
The large central feature appears to deviate from the expected trend, but is in fact due to photon number exceeding the dynamic range of the photon meter.
To explore this region further, we extended the linear range to $\sim$ 0--20 photons using a photon meter with a non-centred refocussing pulse [Fig.~\ref{fig:PhotonNumberChevrons}(d)] and simulated up to 90 Trotter steps ($gt=7.0\pi$), allowing photon oscillations beyond 1.5~\us\ to be observed.
This range operated at the limits of approximately uniform driving given the bandwidth of the 12 \ns\ ($4\sigma$) \QW\ pulses.
At $\gomratio\gtrsim 2$, the photon dynamics in Figs~\ref{fig:PhotonNumberChevrons}(c) and (e) are clearly skewed, causing the observed oscillations to deviate from the expected revival period $\Trabi$ (also observable in the photon parity [Fig.~\ref{fig:QubitVsParity}(g)].
This results from a residual Kerr nonlinearity in $\RR$ inherited from the dispersively coupled ancilla qubit~\cite{Bourassa12}.

Exploring the resonator oscillations more quantitatively, the maximum photon number in each vertical (constant-$\gomratio$) slice [Fig.~\ref{fig:PhotonNumberChevrons}(f)] compares well with the expected ideal behaviour.
The discrepancy between the two curves in the overlapping region results from bandwidth limitations in the high-dynamic-range (HDR) photon meter and the limits in linearity of the number-to-probability mapping for $\QW$.
Because of the sinusoidal conversion, the calibrated value at either end of the range compressed slightly towards the centre from the real photon number.
The measurement saturates at the highest $\gomratio$ even for the HDR meter, suggesting that we observe more than 30 photons (average) building up in the resonator for the strongest USC regions.
Given the Poissonian statistics expected for coherent states, this accesses a resonator subspace of dimension $\sim40$.

\begin{figure*}[t]
  \centering
  \includegraphics[width=\linewidth]{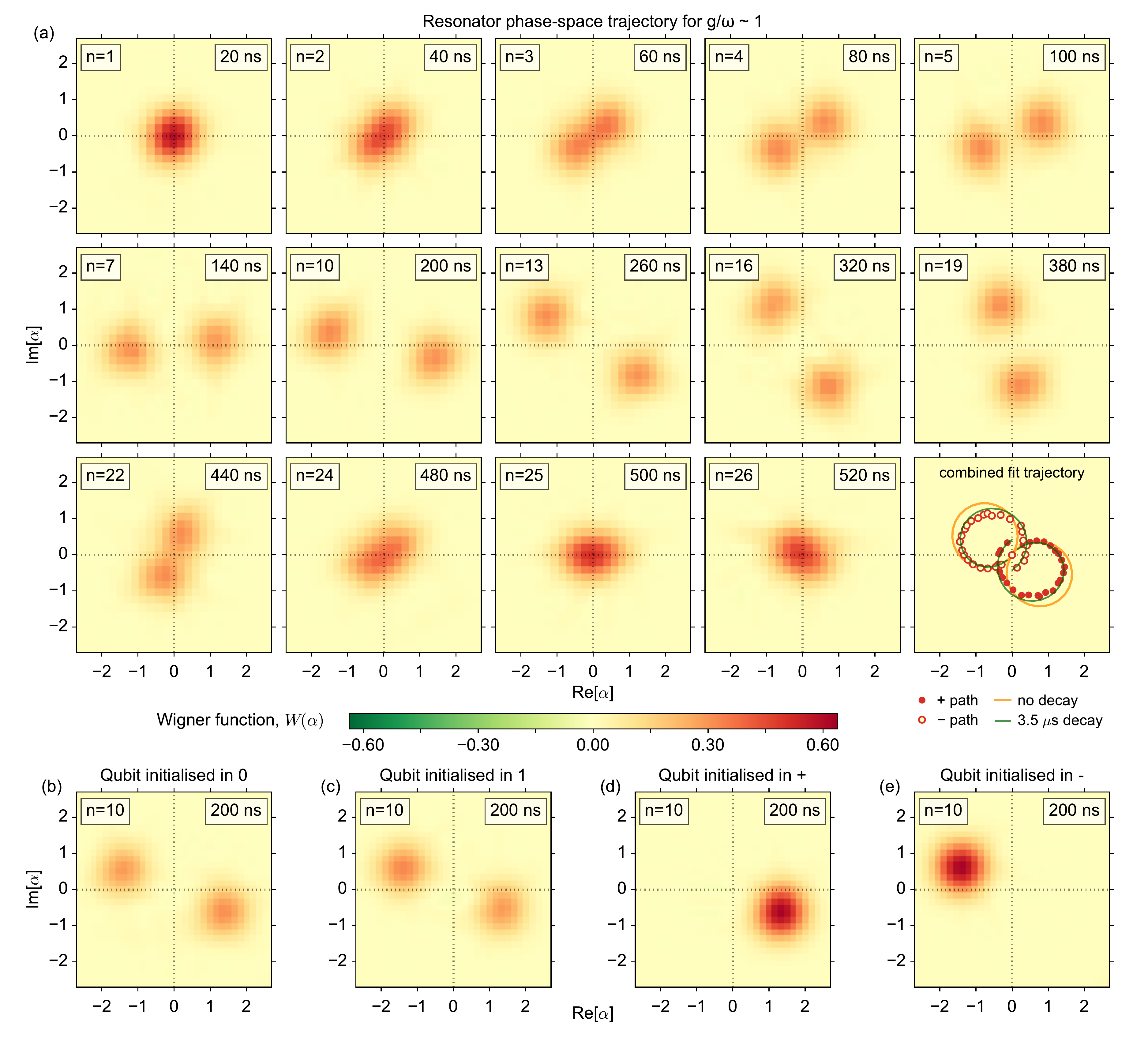}
  \caption{
  {\bf Photon dynamics in phase space in the dUSC regime from maximum-likelihood Wigner tomography.}
  (a) Selected frames from a ``movie'' (measured over $\sim$ 40 hours) showing the phase-space evolution of the resonator reduced state for $\gomratio\sim0.9$, with the final panel showing the full trajectories determined from 2D double-Gaussian fits to the raw data (the full movie is provided in the supplemental material~\cite{SOM}).
  Plotted tomograms are maximum-likelihood reconstructions of direct Wigner tomography measured data with a systematic phase correction (see Methods).
  When the effective drive on the intracavity field created by the Rabi interaction has a strength comparable to the resonator's natural frequency (i.e., $\gR \sim \omrR$), this drive is able to create a significant displacement of the cavity field before the phase-space rotation caused by $\omrR$ brings the field back towards the origin.
  This effect is observed clearly here in the creation of two well-resolved, rotating peaks and subsequent re-coalescence which are characteristic signatures of dUSC dynamics.
  Deviation from the ideal circular trajectories (orange curves) arises from photon decay.
  The measured trajectory shows excellent agreement with a numerical Trotter simulation at $\gR/2\pi=1.79$~\MHz\ which includes resonator $\Toneres=3.5$~\us~(green curves).
  From the fits, we calculate an estimated Wigner function width $\sigma=0.526\pm0.003$, instead of the predicted 0.5, indicating a displacement calibration error of $\sim5\%$~\cite{SOM}.
  Background noise arises from phase instability of microwave sources and frequency stability of the Wigner qubit over the long measurement.
  (b-e) Conditional phase-space evolution illustrated by the resonator Wigner function for different initial states of \QR: (b) $\ket{0}$, (c) $\ket{1}$, (d) $\ket{+}$ and (e) $\ket{-}$.  The phase-space trajectory of \RR\ depends on the qubit state in the $\sigx$ basis, consistent with creation of Bell-cat hybrid entanglement between $\QR$ and $\RR$ of the form: $\ket{+}_{\rm Q}\ket{{+}\alpha}_{\rm R} + \ket{-}_{\rm Q}\ket{{-}\alpha}_{\rm R}$.
  }
  \label{fig:ResoDynamics}
\end{figure*}

Combining the parity measurement with coherent displacements from an external drive allows observation of resonator phase-space dynamics using direct Wigner tomography~\cite{Bertet02, Vlastakis13}.
Figure~\ref{fig:ResoDynamics}(a) shows unconditional maximum-likelihood tomograms (ignoring the state of $\QR$; see Methods) measured after each Trotter step with $\gomratio\sim0.9$ (full movie available~\cite{SOM}), with the full trajectory obtained from two-dimensional double-Gaussian fits of the raw data.
The resonator state displays the clear signatures of USC dynamics, first separating into two distinct Gaussian (coherent-state) peaks which follow opposite circular trajectories before re-coalescing at the origin.
The peaks do not return perfectly to the origin because of photon decay, in agreement with a numerical simulation at $\gR/2\pi=1.79$ which includes $\Toneres=3.5$~\us\ (green curves).

\begin{figure}[t]
  \centering
  \includegraphics[width=\linewidth]{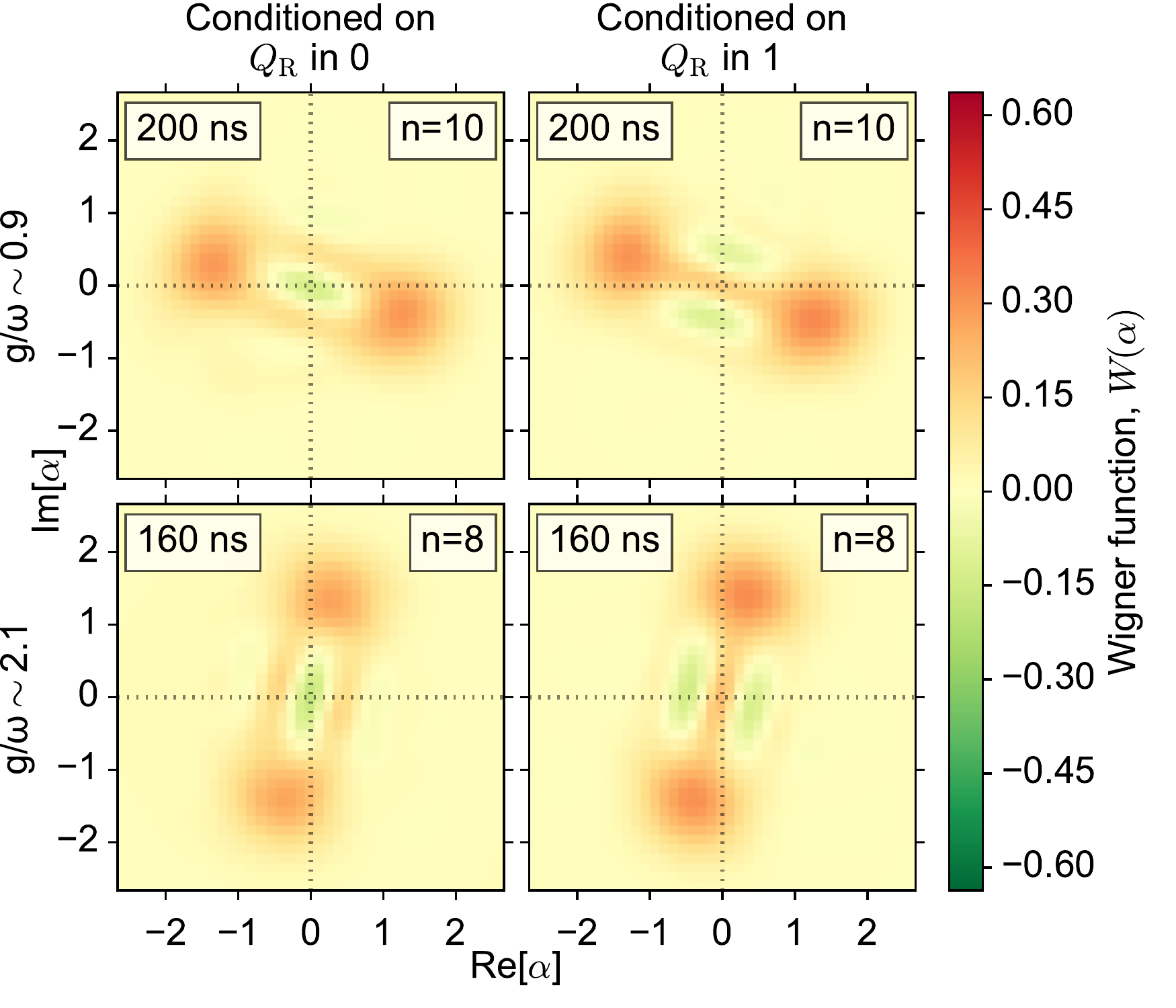}
  \caption{
  {\bf Nonclassical Schr\"odinger cat states of the Rabi resonator from conditioned dUSC-driven entanglement.}
  The plots show resonator Wigner functions from maximum-likelihood state reconstructions for two different ultrastrong coupling strengths with $\gR/\omrR\sim0.9$ (top, 10 Trotter steps) and $\gR/\omrR\sim2.1$ (bottom, 8 Trotter steps), conditioned on measuring $\QR$ in $\ket{0}$ (left) and $\ket{1}$ (right).  The regions of negativity and visibility of several fringes between the well-resolved coherent state peaks are clear signatures of nonclassicality in the Rabi field mode and demonstrates the coherence and entanglement of the underlying qubit-resonator state.  Combined with the qubit conditioning shown in Fig.~\ref{fig:ResoDynamics}, observing clear cat states for both outcomes of the $\QR$ measurement is a clear signature of coherent USC dynamics.}
  \label{fig:ResoCats}
\end{figure}

By capturing the complete resonator quantum state, the Wigner function also enabled the demonstration of coherence in dUSC dynamics, by contrast with photon parity and number measurements, which were largely insensitive to coherence.
Observing this requires correlating the resonator and qubit states, because the coherence is stored in entanglement.
We did this in two ways.
First, we measured the Wigner function after 10 Trotter steps for $\gomratio\sim0.9$ with $\QR$ initialised in states $\ket{0}$, $\ket{1}$, $\ket{+}$ and $\ket{-}$ [Figs~\ref{fig:ResoDynamics}(b--e)].
This showed that the resonator and qubit were correlated, consistent with the expected Bell-cat entanglement.
Second, we ran the simulation for $\gomratio\sim0.9$ and $2.1$ (8 Trotter steps) with the qubit prepared in the excited state, conditioning the $\QW$ measurement on the state of $\QR$ in the $\sigz$ basis (Fig.~\ref{fig:ResoCats}).
For the expected Bell-cat state, an outcome of 0 (1) for $\QR$ leaves the resonator in an odd (even) Schr\"odinger cat state ($\ket{\alpha}\mp\ket{{-}\alpha}$).
Numerical modelling shows that only in the USC regime is negativity in the Wigner function observed for both $\QR$ measurement outcomes.
The negative regions observed in all the Wigner functions demonstrate nonclassicality for all resonator cat states, which arises from coherence in the underlying Bell-cat entanglement.
Reduced visibility is again caused primarily by photon decay, but also by single-shot readout infidelity (here, $\sim$ 85--90\%) and experimental drift over the long measurements.
These different measurements provide clear evidence of qubit-resonator entanglement arising from coherent dUSC dynamics.

\begin{figure*}[t]
  \centering
  \includegraphics[width=\linewidth]{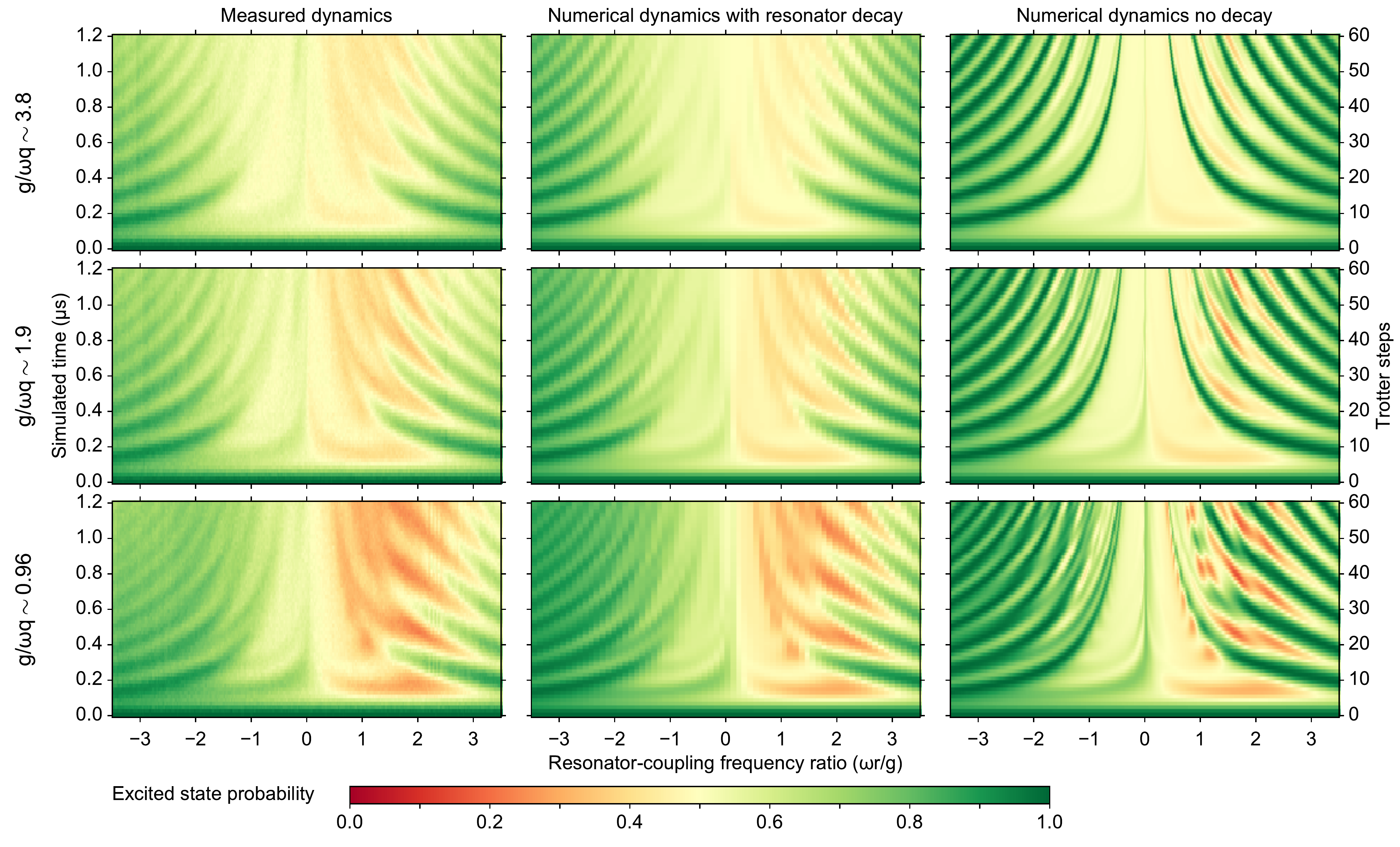}
  \caption{
  {\bf Measured and numerical quantum Rabi model qubit dynamics for nondegenerate qubit frequency.}
  The cases implemented are $\gR/\omqR\sim4$ (top), $\sim2$ (middle) and $\sim1$ (bottom), with the plots showing measured qubit dynamics (left), numerically simulated dynamics of a Trotterised QRM with the measured $\Toneres\sim3.5$~\us\ included (centre), and ideal Rabi dynamics (right).
  The results illustrate that the nondegenerate-qubit dynamics do not deviate significantly from the degenerate-qubit case in the regime where $\omrR \gg \omqR$.
  The measured dynamics exhibit many qualitative features in good agreement with the ideal QRM and show excellent agreement with the numerical Trotter simulation with decay, indicating that the fidelity of the measured results to the ideal case is limited primarily by resonator decay.
  }
  \label{fig:NontrivialRabi}
\end{figure*}

Finally, by detuning the qubit frequency during the JC half of the Trotter steps~[Fig.~\ref{fig:QubitVsParity}(c)], we also experimentally simulate dynamics for the nondegenerate-qubit case of the QRM for effective qubit frequencies $\gR/\omqR \sim$ 4, 2 and 1~(Fig.~\ref{fig:NontrivialRabi}).
These regimes access the full complexity of QRM dynamics.
Comparison with numerical modelling of the ideal QRM (no decoherence) shows that the experimental simulations capture many features of the ideal dynamics, even up to $\gomratio\gg1$.
However, the main deviation from the degenerate-qubit case occurs primarily in the regime $\omrR \lesssim \omqR$~\cite{Albert11}, making the measurements here even more susceptible to photon decay.
Numerical modelling of the digital QRM simulation including the measured $\Toneres$ confirms that simulation fidelity is primarily limited by decay.

Demonstrating stabilisation by decreasing step sizes will be an important part of validating the behaviour of future complex digital simulators achieving quantum advantage~\cite{Cirac12}.
In the supplementary material, we showed that using second-order Trotterisation and decreasing the Trotter step size both significantly improved performance~\cite{SOM}.
This indicates that the simulation is not limited by an error-per-gate noise floor as in previous circuit QED simulations~\cite{Barends15}, and enables us to linearly increase the number of Trotter steps for increasing simulated time, rather than keeping the number fixed~\cite{Barends15, Salathe15, OMalley16}.
This is a crucial step towards the quadratic scaling needed for universal quantum simulation~\cite{Lloyd96}.
In combination, these achievements advance the state of the art in solid-state digital quantum simulation, bringing circuit QED simulators to a level previously attained only in trapped-ion systems~\cite{Lanyon11}.

A QRM simulator has direct advantages over natural USC systems.
Although USC can lead to ground-state entanglement and significant ground-state photon populations, these potentially interesting ground states are not readily accessible in natural USC systems~\cite{Lolli15, Jaako16, Andersen16} without the ability to rapidly (nonadiabatically) tune or switch off the ultrastrong coupling.
In systems where the coupling reaches many gigahertz, tuning system parameters on this timescale represents a significant technical challenge~\cite{Forn-Diaz16, Yoshihara16}.
In our simulator, however, cavity photons are always real (not virtual), detectable and usable, and it is straightforward to nonadiabatically tune system parameters to implement quantum quenches~\cite{Zurek85}.
This makes a circuit QED chip with natural JC interactions an ideal platform to explore the preparation of interesting ground states in future experiments.
The challenge is that the simulator decay processes differ from those in a natural USC system and do not move the system towards the USC ground state~\cite{Beaudoin11}.
This highlights the need to improve $\Toneres$ so that photon decay does not limit the dynamics.
It should be possible to improve $\Toneres$ ten-fold using novel processing methods~\cite{Bruno15}.

Finally, the phase technique we have developed to define a rotating frame via single-qubit pulses introduces a precise and flexible paradigm for engineering artificial Hamiltonians which can be applied across architectures such as trapped ions and cold atoms~\cite{Lanyon11, Pedernales15, Felicetti16}.
In combination with the number of Trotter steps demonstrated, the technique will allow accurate simulation of the time-dependent Hamiltonians~\cite{Lanyon11, Barends15, Barends16} required to perform adiabatic preparation of USC ground states.
It is therefore ideally suited for exploring novel quantum phase transitions relying on extreme coupling regimes recently identified for the QRM~\cite{Felicetti16, Hwang15, Puebla16}.
Further, by extending to small-scale Dicke-model systems~\cite{Mezzacapo14, Lamata16}, it will avoid the problem of additional nonlinear evolution terms~\cite{Lamata16} which have been suggested to prevent the onset of a long-predicted superradiant phase transition in a range of physical systems~\cite{Nataf10, Viehmann11, DeLiberato14, Jaako16}.

\renewcommand{\theequation}{M\arabic{equation}}
\renewcommand{\thefigure}{M\arabic{figure}}
\renewcommand{\thetable}{M\arabic{table}}
\setcounter{figure}{0}
\setcounter{equation}{0}
\setcounter{table}{0}

\section{Methods}

\subsection{Phase-controlled Trotterisation of the quantum Rabi model}

In the digital QRM simulation proposed in Ref.~\cite{Mezzacapo14}, the effective parameters of the simulated Rabi Hamiltonian are $\gR=g$, $\omrR=2\Delr$ and $\omqR=\Delq^{\rm JC}-\Delq^{\rm AJC}$, where $\Delr = \omr-\omrf$ and $\Delq=\omq-\omrf$ are defined relative to a rotating frame.
This rotating frame is essential to reaching the USC regime with weakly anharmonic transmon qubits, by allowing us to tune the simulated $\omrR$ and $\omqR$.
Typically, the frequency of a rotating frame is set by a physical generator or drive signal that defines a rotation or a measurement basis.
In the digital simulation, the rotating frame is still abstract, since no drive is used to induce an interaction.
Here, we describe a method we have developed for controlling the frequency of the rotating frame which is simple, high-resolution and flexible.

In the experiment, the frequency of the rotating frame is defined by the rotation axes of the bit-flip pulses (set by the pulse phase) that convert every second JC interaction into an effective AJC interaction.
The qubit, however, is driven at its bottom sweet spot, around 1~\GHz\ below the resonator.
The drive generator phase therefore changes rapidly by comparison with the target rotating frame and it is necessary to reset the generator back in phase with the target rotating frame each time a pulse is applied.

We now derive the relation between the bit-flip pulse phases and the rotating-frame frequency.
The symmetric, second-order Trotter step for the digital QRM simulation is:
\begin{align}
\UTrRabi\br{\tau} &= \Ujc^\frac{1}{2}\br{\tau} \Uajc\br{\tau} \Ujc^\frac{1}{2}\br{\tau},
\end{align}
where $\Ujc\br{\tau}=\exp(-i\Hjc \tau/\hbar)$ and an arbitrary AJC step
\begin{align}
\Uajc\br{\tau} &= \Rgen{\phi_1}{\pi} \exp\br{\frac{-i \Hjc \tau}{\hbar}} \Rgen{\phi_2}{\pi},
\label{eq:AJCStepBitFlips}
\end{align}
is defined by the phases used to set the rotation axes $\phi_{1,2}$ of the bit flips $\Rgen{\phi}{\pi}$.
Writing the JC Hamiltonian in the rotating frame of the resonator, and using the identity $\Rgen{\phi}{\pi} = \Rgen{z}{\phi} \Rgen{x}{\pi} \Rgen{z}{-\phi}$, gives:
\begin{widetext}
\begin{align}
\Uajc\br{\tau} &= \Rgen{z}{2\phi_1} \sigx \exp\br{-i\epsilon \br{a \sigp {+} \ad\sigm}} \sigx \Rgen{z}{-2\phi_1}, \\
&=\exp\br{i\dphi\sigz/2} \exp\br{-i\phi_\Sigma\sigz/2} \exp\br{-i\epsilon \br{a \sigm {+} \ad\sigp}} \exp\br{i\phi_\Sigma\sigz/2} \exp\br{i\dphi\sigz/2}, \\
&= \exp\br{i\dphi\sigz/2} \exp\br{-i\epsilon \br{a \sigm e^{-i\phi_\Sigma} {+} \ad\sigp e^{i\phi_\Sigma}}} \exp\br{i\dphi\sigz/2},
\label{eq:AJCStepBitFlipsRewrite}
\end{align}
\end{widetext}
where $\epsilon = g\tau$, $\phi_\Sigma=\phi_1+\phi_2$, $\dphi=\phi_2-\phi_1$ and we have set $\Delqr^{\rm AJC}=0$.
Equation~(\ref{eq:AJCStepBitFlipsRewrite}) is reached by noting that $e^{-i\phi_\Sigma\sigz/2}\sigpm e^{i\phi_\Sigma\sigz/2}=\sigpm e^{\pm i\phi_\Sigma}$.

Next, noting that $\dphi = \pi \omrR \tau \ll1$ if $\tau \ll 1/\omrR$, and providing the Trotter condition $\epsilon = g\tau \ll 1$ is fulfilled, we can combine exponentials using Trotter approximations to give first:
\begin{align}
\Uajc\br{\tau} \approx \exp\br{i\dphi \sigz - i\epsilon \br{a \sigm e^{-i\phi_\Sigma} {+} \ad\sigp e^{i\phi_\Sigma}}},
\end{align}
and then the full Trotter step
\begin{widetext}
\begin{align}
\UTrRabi\br{\tau} &\approx
\exp\left[ i \br{2\dphi+\Delqr^{\rm JC}\tau} \frac{\sigz}{2}
- i\epsilon \br{a \sigp + \ad\sigm + a \sigm e^{-i\phi_\Sigma} + \ad\sigp e^{i\phi_\Sigma}}\right].
\end{align}
\end{widetext}
So far, we have considered arbitrary $\phi_1$ and $\phi_2$.
In the experiment, however, we keep $\dphi$ constant for all sequential pairs of bit flips.
Specifically, for the $n$th Trotter step, the two phases are $\phi_1 = \phi_0 + \br{2n{-}2}\dphi$ and $\phi_2 = \phi_0 + \br{2n{-}1}\dphi$, where the choice of $\phi_0$ has no effect on the dynamics.
Setting $\phi_0=3\dphi/2$ gives $\phi_\Sigma = 4n\dphi$, and the $n$th Trotter step can be rewritten in terms of a frequency $\omega_0=2\dphi/\tau$ and a time $t_n=n\tau$:
\begin{widetext}
\begin{align}
\Urabi^{(n)}\br{\tau} &=
\exp\left[ i \br{\omega_0+\Delqr^{\rm JC}}\tau \frac{\sigz}{2}
- i\epsilon \br{a \sigp + \ad\sigm + a \sigm e^{-i2\omega_0 t_n} + \ad\sigp e^{i2\omega_0 t_n}}\right].
\end{align}
which corresponds to an effective Hamiltonian:
\begin{align}
\frac{H_{\rm eff}}{\hbar} \equiv \frac{-\overline{H}_{\rm eff}}{\hbar}
= -\br{\omega_0 +\Delqr^{\rm JC}}\frac{\sigz}{2}
+ g \br{a \sigp + \ad\sigm + a \sigm e^{-i2\omega_0 t} + \ad\sigp e^{i2\omega_0 t}}.
\end{align}
\end{widetext}
Until this point, the calculation has been carried out with both qubit and resonator in a frame rotating with the resonator.
We now transform $\overline{H}_{\rm eff}$ into a rotating frame where both qubit and resonator are rotating at frequency $\br{-\omega_0}$, i.e., with $H_0=-\hbar\omega_0 \br{-\sigz/2+\ada}$, giving:
\begin{align}
\nonumber
&\exp\br{iH_0t/\hbar} \frac{\overline{H}_{\rm eff}-H_0}{\hbar} \exp\br{-iH_0t/\hbar} \\
&= \Delqr^{\rm JC} \frac{\sigz}{2} + \omega_0 \ada
- g \br{a \sigp + \ad\sigm + a \sigm + \ad\sigp}.
\end{align}
Thus, in the new rotating frame, the final effective Hamiltonian implemented by the Trotterisation is:
\begin{align}
\frac{H_{\rm eff}}{\hbar} = -\Delqr^{\rm JC} \frac{\sigz}{2} - \omega_0 \ada
+ g \br{a + \ad}\br{\sigp + \sigm}.
\end{align}
This completes the mapping of the phase-controlled Trotterisation into the form of a simulated Rabi Hamiltonian and we can now identify the effective simulated parameters $\gR=g$, $\omqR=\Delqr^{\rm JC}$ and $\omrR=-\omega_0=-2\dphi/\tau$.
It is worth emphasising here that, by controlling the phase difference between successive bit-flip pulses on the \emph{qubit}, we are able to define the rotating frame frequency $\omega_0$, and hence the effective \emph{resonator} frequency in the simulated Hamiltonian.

\subsection{Trotter step}

For a second-order Trotter step with simulated time $\tau$, the Trotter step consists of 3 flux pulses ($\tau/2$, $\tau$ and $\tau/2$) and 2 single-qubit rotations with buffers separating the different gates.
Adjacent $\tau/2$ flux pulses from neighbouring Trotter steps are implemented as a single flux pulse of length $\tau$.
Each flux pulse was followed by a 5~\ns\ phase-compensation flux pulse~\cite{SOM}.
For most of the data presented in this work, the simulated $\tau=20$~\ns.
The qubit drive pulses on $\QR$ were 16~\ns\ total duration ($4\sigma$) and the pulses buffers were 10~\ns.
The total Trotter step for $\tau=20$~\ns\ was therefore $\tau_{\rm step}=122$~\ns.
In addition to the drive-pulse phase advance required to define $\omrR$, another linear phase advance $\Delta\phi=(\omq^{\rm drive}-\omr)\tau_{\rm step}/2$ is required to compensate the rapid rotation of the qubit drive with respect to the resonator frequency.

\subsection{Qubit control}

Qubit rotations were implemented using DRAG pulses~\cite{Motzoi09, Chow10b}, with a Gaussian envelope in the $X$ quadrature and a derivative-of-Gaussian envelope in the $Y$ quadrature.
The $4\sigma$ pulse durations were 16~\ns\ for $\QR$ and 12~\ns\ for $\QW$.
The performance of the Trotter sequences, which contained up to 180 bit-flip pulses, was very sensitive to details of the $\QR$ pulse calibrations.
In particular, the drive amplitude was calibrated using a sequence of 50 $\pi$-pulse pairs preceding a single $\pi/2$ pulse.
All parameters were typically calibrated just before launching a long measurement.
The drive amplitude was intermittently recalibrated during the scans.
Because only 2 or 3 pulses were applied to $\QW$ for the photon measurements, it was optimised using the AllXY sequence~\cite{thesisReed13} of 21 combinations of two $\sigx$ and $\sigy$ rotations (either $\pi/2$ or $\pi$).
The frequency of $\QW$ was regularly calibrated during photon measurements using Ramsey sequences.

\subsection{Wigner tomography reconstructions}

Tomograms shown in Figs~\ref{fig:ResoDynamics} and~\ref{fig:ResoCats} are maximum likelihood reconstructions~\cite{Hradil97, James01} of the resonator quantum state from direct Wigner tomography measurements~\cite{Vlastakis13}.  The Wigner function at a phase-space position $\alpha$ is:
\begin{align}
W\br{\alpha} = \frac{2}{\pi} \tr\sqbr{\Pi D^\dag\br{\alpha} \rho_{\rm res} D\br{\alpha}}
=  \frac{2}{\pi} \tr\sqbr{M_\alpha \rho_{\rm r}},
\end{align}
where $\rho_{\rm r}$ is the resonator density matrix, $\Pi = \sum_n ({-}1)^n \ket{n}\bra{n}$ is the photon parity operator and $D\br{\alpha}$ is the coherent displacement operator.
For each measured $\alpha$, we calculated $M_\alpha = D\br{\alpha} \Pi D^\dag\br{\alpha}$ using an operator dimension much larger than the largest $|\alpha|^2$ in the measured phase space, to avoid edge effects when calculating $D\br{\alpha}$.
The $M_\alpha$ were then truncated to a maximum photon number sufficient to capture all of the reconstructed state, but small enough to allow fast reconstructions and ensure an informationally complete set of operators ($n_{\rm max}=12$ and 8 for tomograms in Figs~\ref{fig:ResoDynamics} and~\ref{fig:ResoCats}, respectively).
The maximum likelihood reconstruction was carried out using convex optimisation~\cite{Boyd04, Langford13}.
In Fig.~\ref{fig:ResoDynamics}, a systematic phase correction was applied to the density matrices to correct for a miscalibration of the resonator drive phase used in the coherent displacement.
Finally, the reconstructed density matrix was then used to calculate the plotted Wigner functions.

\section{Acknowledgements}

We acknowledge experimental contributions from R.~N.~Schouten, O.~P.~Saira and C.~C.~Bultink, software developments by M.~A.~Rol, S.~Asaad and G.~de Lange, and discussions with 
G.~Kirchmair, U.~Las Heras, A.~Mezzacapo, L.~Lamata, E.~Solano, W.~J.~Munro, C.~Ciuti, and M.~J.~Hartmann.
This research was supported by the EU project ScaleQIT, the ERC Synergy grant QC-lab, the Netherlands Organisation for Scientific Research as part of the Frontiers of Nanoscience program (NWO/OCW) and a Vidi Grant (639.042.423), the Dutch Organization for Fundamental Research on Matter (FOM), and Microsoft Corporation Station Q.

\section{Author contributions}

N.K.L. designed and fabricated the device, with input from M.K., A.B., C.D., F.L.\ and L.D.C.
A.B., D.J.T.\ and A.E.\ sputtered the NbTiN thin film.
N.K.L.\ and R.S.\ performed measurements and data analysis, with contributions from C.D.\ and F.L.
N.K.L., M.K.\ and L.D.C.\ carried out numerical modelling.
L.D.C., N.K.L.\ and M.K.\ developed the phase-based Trotterisation.
N.K.L.\ wrote the manuscript, with input from all coauthors.
L.D.C.\ supervised the project.

\section{Author information}

The authors declare that they have no competing financial interests.
Correspondence and requests for materials should be addressed to L.D.C. (l.dicarlo@tudelft.nl).


%

\clearpage

\renewcommand{\theequation}{S\arabic{equation}}
\renewcommand{\thefigure}{S\arabic{figure}}
\renewcommand{\thetable}{S\arabic{table}}
\setcounter{figure}{0}
\setcounter{equation}{0}
\setcounter{table}{0}

\section{Supplementary Information}

This supplement provides experimental details and additional data supporting the claims in the main text.

\subsection{Experimental setup}

\begin{figure*}[t]
  \centering
  \includegraphics[width=\linewidth]{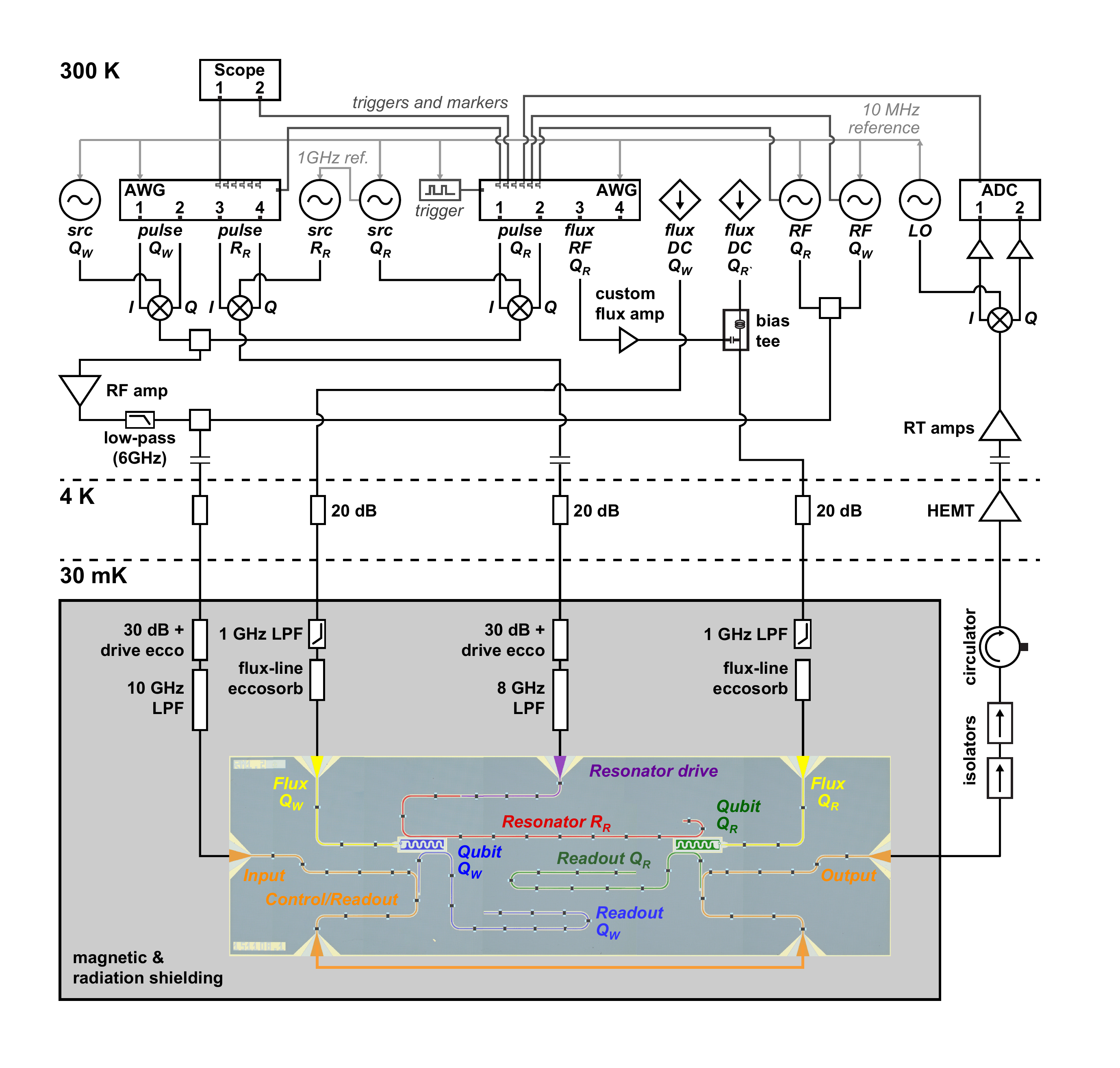}
  \caption{
  Experimental schematic showing the connectivity of microwave electronics and components in and outside the dilution refrigerator.
  The sample mounted below the mixing chamber typically remained at around 30~\mK.
  Qubit and resonator drive lines and flux-bias lines were thermalised and attenuated at the 4-$\K$ and 30-$\mK$ stages and were low-pass filtered before arriving at the sample.
  The qubits and resonator drive pulses were generated by AWGs and IQ mixers.
  Home-built low-noise current sources provided DC bias currents for qubit frequency tuning, which were combined with fast frequency-tuning bias pulses using reactive bias tees.
  AWG markers provided the gating for pulse-modulated measurement pulses.
  }
  \label{fig:ExperimentalSchematic}
\end{figure*}

The sample and low-temperature microwave components were mounted inside magnetic and infrared radiation shielding consisting of two layers of cryogenic mu metal around a layer of aluminium, with an internal layer of copper foil coated in a mixture of silicon carbide and Stycast (2850 FT)~\cite{Barends11SOM}.
Microwave coaxial cables are connected to the PCB-mounted chip via non-magnetic SMP connectors (Rosenberger).

The qubit drive and read-out tones are sent through two dedicated feedlines which are connected via a short coaxial cable off-chip.
The input line for the qubit drives is filtered at the mixing chamber with 30~$\dB$ cold attenuation, a small home-built inline eccosorb filter and a 10~$\GHz$ low-pass filter (K\&L 6L250-10000/T20000-0/0).
[The resonator input line filter is 8~$\GHz$ low-pass (K\&L 6L250-8000/T18000-0/0).]
The output line passes through two 3--12~$\GHz$ isolators (Pamtech CWJ1019K) and a circulator (Quinstar CTH0408KCS) mounted above the mixing chamber on the way to a 4--8~\GHz\ cryogenic HEMT amplifier (Low-Noise Factory LNF-LNC4\_8A), two room-temperature amplifiers (Miteq AFS3-04000800-10-ULN, then AFS3-00101200-35-ULN-R), RF demodulation (Marki 0618LXP IQ mixer) and amplification, and finally digitised in a data acquisition card (AlazarTech ATS9870).
The flux-bias lines are filtered at the mixing chamber with 1.35~\GHz\ low-pass filters (Minicircuits VLFX-1350) followed by home-built eccosorb filters.
All input lines are thermalised with 20~$\dB$ attenuators mounted at the 4~$\K$ plate.  The microwave input lines and output line are connected to the fridge through a DC block.

Qubit and resonator drive pulses are created via single-sideband modulation with IQ mixers and generated by two arbitrary waveform generators (AWGs; Tektronix AWG5014).
We use a 3--7~$\GHz$ IQ mixer (Marki 0307MXP) for the resonator and two custom-built 4--8.5~$\GHz$ IQ mixers (QuTech F1c: DC--3.5~$\GHz$ IF bandwidth) for the qubit drives.
The qubit drive pulses were amplified by a high-power (35~\dB) microwave amplifier (Minicircuits ZV-3W-183) before passing through a 5.5~$\GHz$ low-pass filter (Minicircuits LFCN 5500+) to minimise amplifier noise at the readout resonator frequencies.

Most microwave units receive a 10~$\MHz$ reference from a microwave generator (Agilent E8257D) via a home-built distribution unit.
However, the generators used for driving $\QR$ and $\RR$ (R\&S SGS100A) synchronised directly via a 1~$\GHz$ reference.
This was critical to achieving the phase stability required to measure $\RR$ Wigner functions during measurement runs lasting up to 40 hours.
The frequencies for these two generators were also always set to a multiple of the trigger repetition rate (5~\kHz), to ensure a stable phase relationship.
For phase-sensitive measurements, a 500~$\MHz$ scope (Rigol DS4034) monitored the relative trigger timing between the master and slave AWGs to select consistent delay configurations between the AWG outputs.

Home-built low-noise current sources mounted in a TU Delft IVVI-DAC2 rack provided precision DC bias currents for flux tuning of the qubit frequencies.
The DC bias for $\QR$ was combined with the amplified output of one channel of the master AWG (the same as used for generating $\QR$ drive pulses) using a reactive bias tee (Minicircuits ZFBT-6GW+).
The flux pulses from the AWG were amplified using a home-built 2 V/V flux-pulse amplifier.

\subsection{Device fabrication}

\begin{figure}[t]
  \centering
  \includegraphics[width=0.95\linewidth]{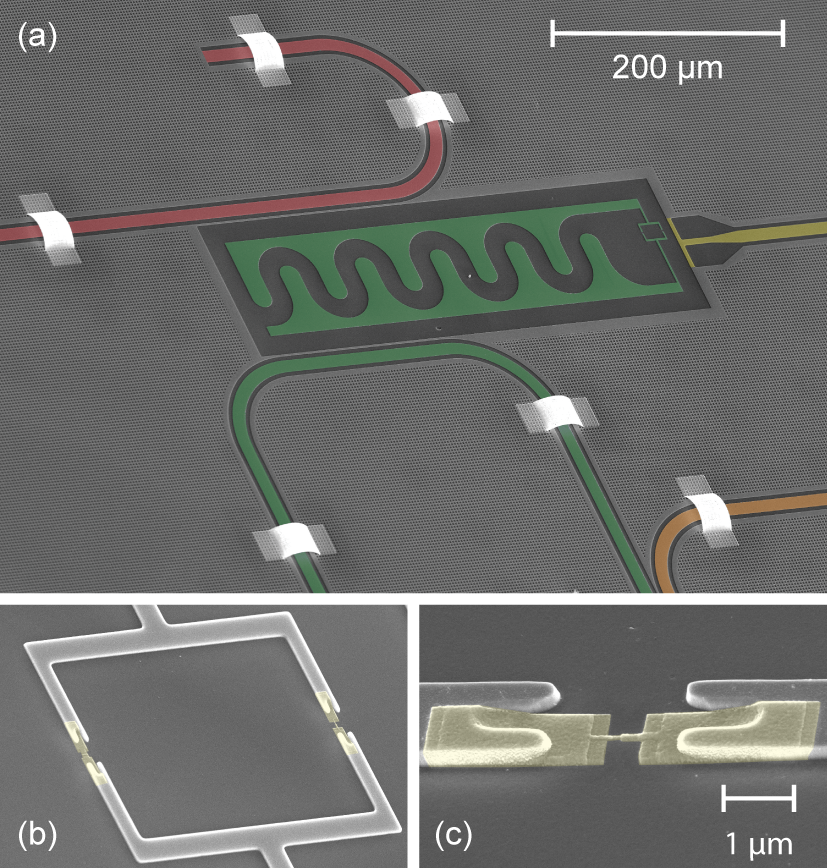}
  \caption{
  SEM images of a sister device with added false colour.
  (a) Rabi qubit ($\QR$) with coupling to the Rabi resonator ($\RR$, above) and readout resonator (below), showing the centred flux-bias line and displaced SQUID loop.  $\QR$ is coupled to $\RR$ near its shorted end in order to achieve the required small coupling $g$.  (b, c) Josephson junctions are contacted to the NbTiN SQUID loop fingers using small bays to achieve better contact.  In (b), it is possible to see the large asymmetry in junction size, with a zoom on the small junction in (c).
  }
  \label{fig:DeviceSEM}
\end{figure}

The device was fabricated using a method similar to that of Ref.~\onlinecite{Asaad16SOM}, but with several specific improvements:
\begin{enumerate}
\item The transmon design includes a rounded spacing between the shunt capacitor plates [Fig.~\ref{fig:DeviceSEM}(a)] to avoid the regions of high electric field which can increase sensitivity to interface two-level fluctuators~\cite{Gambetta16SOM}.
\item The flux-bias line was centred between the transmon capacitor plates to symmetrise the capacitive coupling with the goal of decoupling the qubits from possible decay-inducing effects of voltage noise fluctuations on the flux-bias lines.
\item As in our previous work~\cite{Asaad16SOM}, the transmon qubits were patterned with niobium titanium nitride (NbTiN) capacitor plates to further reduce susceptibility to noise from two-level fluctuators.
Prior to evaporation of the aluminium (Al) junction layers, a short hydrogen-fluoride (HF) dip removed surface oxides to facilitate a good contact between the evaporated Al and NbTiN thin film.
To avoid contact problems caused by unwanted etching into the silicon substrate during patterning of the NbTiN, we: 1) optimised the reactive-ion etch (RIE) recipe and duration to minimise the substrate etch and eliminate underetch (under the NbTiN); and 2) introduced a narrow bay in the NbTiN fingers at the contact point to create a softer etch for more reliable contact [Fig.~\ref{fig:DeviceSEM}(c)].
\item The junction development process and double-angle evaporation parameters were optimised to improve the reliability of the very small junction sizes needed for the asymmetric qubit [Fig.~\ref{fig:DeviceSEM}(b)].
\end{enumerate}

\subsection{Device operating parameters and qubit performance}

\begin{table*}[ht]
\begin{tabular}{p{2.cm} p{2.5cm} p{1.5cm} p{1cm} p{1.0cm} p{1.5cm}}
Component & \multicolumn{2}{c}{Frequency domain} && \multicolumn{2}{c}{Time domain}\\
\hline
\QR & $f_{\rm max}$ & 6.670 \GHz && \multicolumn{2}{l}{At operating point:} \\
& $f_{\rm min}$ & 5.451 \GHz && $\Tone$ & 20--30~\us \\
& $\alpha$ (asymmetry) & 0.68 && $\Techo$ & 30--60~\us \\
& $E_{\rm C}/2\pi$ & $-281$ \MHz && $\Ttwostar$ & 20--50~\us \\
& $f_{\rm readout}$ & 7.026 \GHz &&& \\
& $g_{\rm readout}/2\pi$ & $43$~\MHz &&& \\
\hline
\RR & $f$ & 6.381 \GHz && $\Toneres$ & 3--4~\us \\
& $g_{\rm r}/2\pi$ (to \QR) & $1.92$ \MHz && $g_{\rm r}/2\pi$ & $1.95$ \MHz \\
& $\chi_{\rm w}/\pi$ (to \QW) & $-1.26$~\MHz &&& \\
\hline
\QW & $f_{\rm max}$ & 5.653 \GHz && \multicolumn{2}{l}{At operating point:} \\
& $f_{\rm exp}$ & 5.003~\GHz && $\Tone$ & 30--40~\us \\
& $E_{\rm C}$ & ------ && $\Techo$ & 5--7~\us \\
& $f_{\rm readout}$ & 6.940 \GHz && $\Ttwostar$ & 1.5--1.8~\us \\
& $g_{\rm readout}/2\pi$ & $42$~\MHz && \multicolumn{2}{l}{At top sweet spot:} \\
&&&& $\Techo$ & 30--60~\us \\
&&&& $\Ttwostar$ & 20--50~\us \\
\hline
\end{tabular}
\caption{Measured device parameters and qubit and resonator performance.
The coupling strength between $\QR$ and $\RR$ was measured both by spectroscopy of the avoided crossing, and time-domain measurement of the vacuum Rabi oscillation frequency.
For both qubits, Ramsey sequences measured at the sweet spots exhibited beating consistent with quasiparticle tunnelling~\cite{Riste13SOM}.  $\Ttwostar$s reported here were measured by fitting a decaying double sinusoid to a long, beating Ramsey signal and represents the underlying coherence of the qubits.  At the operating point for $\QW$ far from the sweet spot, no beating was observed in the Ramsey measurements.}
\label{tab:DeviceParameters}
\end{table*}

\begin{figure*}[t]
  \centering
  \includegraphics[width=1.\linewidth]{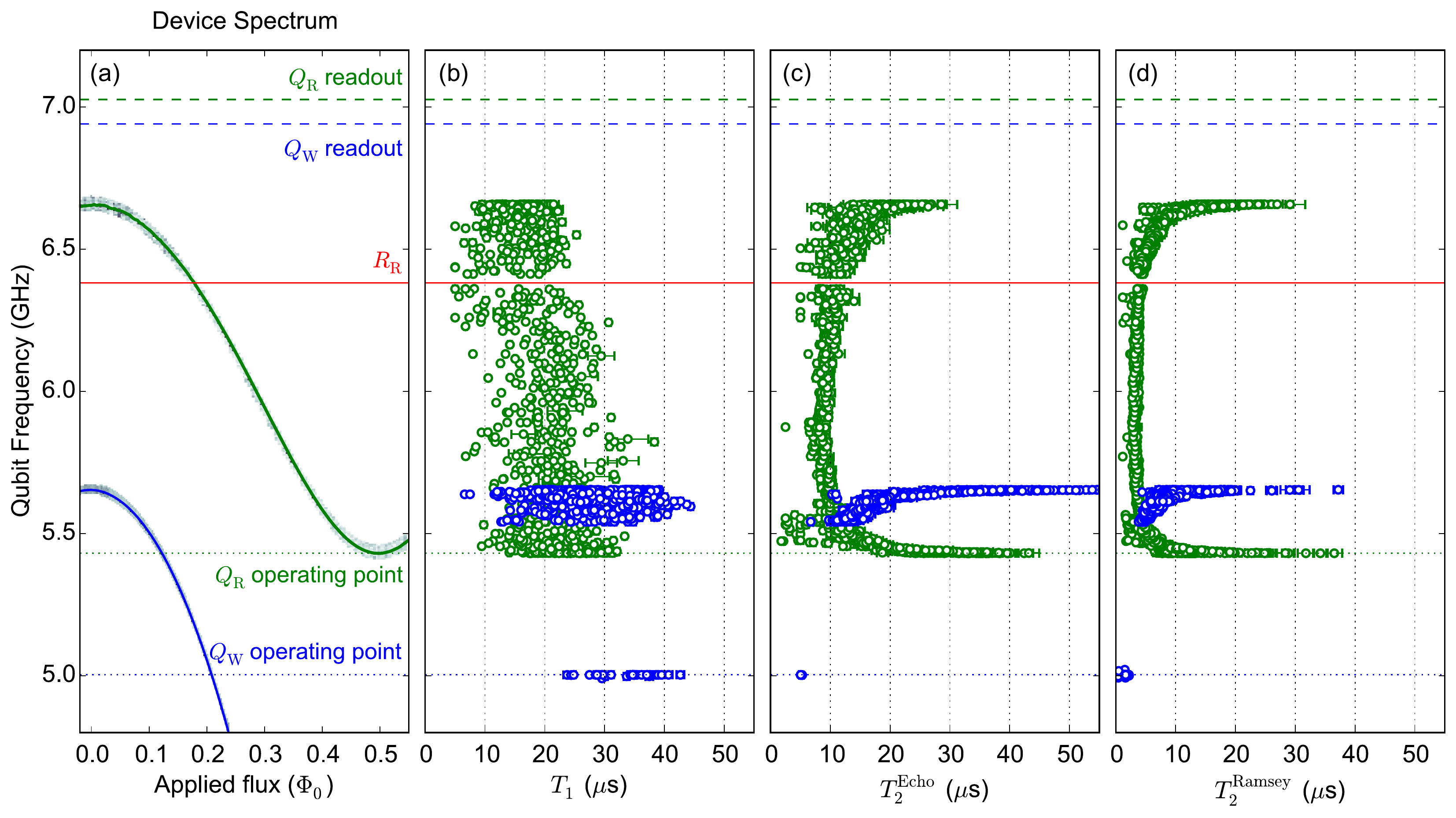}
  \caption{
  Schematic showing measured spectral arrangement for the digital Rabi quantum simulator device and the qubit coherence times.
  (a) Measured data for the 0--1 transition of the Rabi qubit $\QR$ (green curve) and the Wigner qubit $\QW$ (blue curve) are plotted as a function of applied flux in units of $\Phi_0$.
  Also shown are the frequencies of the Rabi resonator $\RR$ (red: $\omr = 6.381 \GHz$) and the readout resonators for $\QR$ (green dashed: $\sim 7.03$ \GHz) and $\QW$ (blue dashed: $\sim 6.94$ \GHz).
  The operating points of the qubits for the Trotter simulation are given by the green and blue dotted lines for $\QR$ and $\QW$, respectively.
  (b, c, d) Time constants measured for $\QR$ (green) and $\QW$ (blue) for (b) $\Tone$, (c) $\Techo$ and (d) $\Ttwostar$.  Note that, at the sweet spots, measured qubit $\Ttwostar$ times here are limited by slow frequency-switching processes in the qubits such as quasiparticle tunnelling~\cite{Riste13SOM}.  
  }
  \label{fig:FrequencyScheme}
\end{figure*}

Figure~\ref{fig:FrequencyScheme}(a) shows the frequencies for the two qubits and three resonators on the device as a function of the applied qubit flux in units of the flux quantum $\Phi_0=h/2e$, along with the operating points for both qubits during the quantum simulation experiments.
Measured device parameters are summarised in Tab.~\ref{tab:DeviceParameters}.
Qubit $\Tone$, $\Techo$ and $\Ttwostar$ decay times are shown as a function of qubit frequency in Fig.~\ref{fig:FrequencyScheme}(b,c,d).

At the operating point, the Rabi qubit $\QR$ was designed to sit below the resonator $\RR$ and be pulsed up into resonance with it to avoid continually crossing the resonator with the $\QR$'s 1--2 transition during the long flux-pulse sequence.
Because of significant protocol times and two operating points, an asymmetric qubit design with two flux-insensitive ``sweet'' spots was used for \QR~\cite{Koch07SOM}, with drive pulses applied at its bottom sweet spot.
The first-order flux insensitivity at this point also mitigated some of the impact of rapid, long-range flux-pulsing on the qubit pulse tuning.
The maximum and minimum frequencies for $\QR$ in the final cooldown were 6.670~\GHz\ and 5.451~\GHz, respectively.

The asymmetric design also minimised the stringent challenge of targetting the qubit frequency to resonator closely on the scale of the very small coupling frequency.
Ideally, the resonator would have been closer to the qubit top sweet spot to maximise phase coherence also during the interaction pulses.
However, with the asymmetric design, the reduced flux gradient relaxes this constraint.
With an asymmetry parameter of $\alpha = (\EJmax{-}\EJmin)/(\EJmax{+}\EJmin) \sim 0.68$, the Ramsey time $\Ttwostar$ for $\QR$ did not typically drop below a few microseconds, even at the positions with steepest flux gradient.

The asymmetry of $\QR$ was smaller than targetted, with the result that the bottom sweet spot was also lower in frequency than intended.
The ancilla qubit $\QW$ (a standard symmetric transmon) was therefore operated around 650~$\MHz$ below its own maximum-frequency sweet spot of 5.653~$\GHz$.
At this operating point, its $\Ttwostar$ was typically $\gtrsim 1.5$~\us.
Because we were able to drive \QW\ and achieve good photon-sensitive operation at this lower position, we chose not to rapidly tune its frequency up to the sweet spot to perform the photon meter measurements.

To identify the flux operating point that positioned $\QR$ precisely at the bottom sweet spot, we applied the following procedure.
We first decoupled the applied DC qubit fluxes, applying the appropriate linear correction to compensate for flux cross-talk.
Then, after positioning $\QW$ roughly at its selected operating point, we applied a simple excitation swapping sequence for $\QR$ with $\RR$ with fixed swap time (near a full swap) and varying amplitudes of positively and negatively directed pulses.
Finally, we varied the applied flux on $\QR$ and identified the operating point as the symmetric flux point where the qubit hit the resonance for positive and negative pulses of equal amplitude.
We were able to identify this point to 1 part in 5000.
Because the precise choice of operating frequency for $\QW$ was not critical, any slight shift in frequency due to residual DC cross-talk remaining after the flux decoupling measurements was unimportant.

\subsection{Calibration of the flux distortions}

\begin{figure*}[t]
  \centering
  \includegraphics[width=\linewidth]{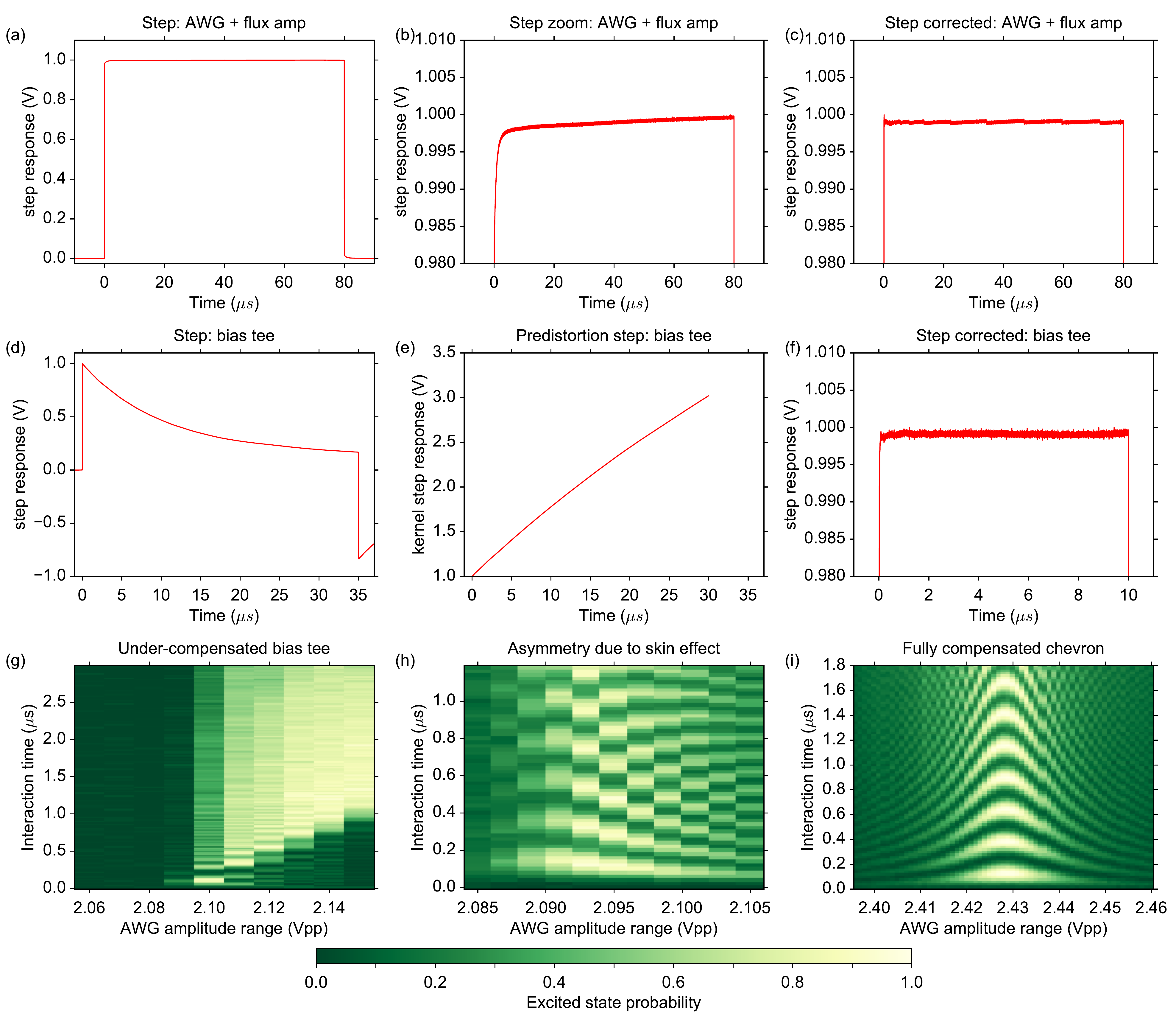}
  \caption{Calibration of the flux distortions.
  (a,b) Step response of the amplified AWG flux channel output with a zoom in (b), measured using a fast oscilloscope.
  (c) Corrected step response achieved using one linear response correction and three exponential decay corrections with parameters ($\tau$, $\alpha$): (5.1 \us, 0.0012), (670 \ns, 0.015) and (520 \ns, -0.00037) (see text for details).
  (d, e) Measured step response (d) and numerically calculated predistortion step response (e) after the bias-tee.
  (f) Corrected step response achieved using a quadratic bias tee correction (see text for details).
  (g) Distorted flux ``chevron'' measured with the corrections applied in (e).
  (h) Dramatically improved chevron obtained after sweeping one parameter in the bias-tee correction (that corresonding to the standard RC time constant).  The asymmetric signature observed here is characteristic of the low-pass filtering effect produced by the skin effect in the coaxial cables.
  (i) A well-compensated chevron obtained after applying a correction for the skin effect and several more exponential decay corrections with ($\tau$, $\alpha$): (350 \ns, -0.0063), (600 \ns, -0.0037), (1500 \ns, -0.002), (100 \ns, -0.0017) and (30 \ns, 0.0036).
}
  \label{fig:FluxCals}
\end{figure*}

Implementing the digital Trotterisation of the Rabi model proposed in Ref.~\cite{Mezzacapo14SOM} required tuning the qubit frequency with a long series of square interaction pulses.
To achieve this, it was necessary to compensate for the filtering effects of electronics and microwave components in the line (Fig.~\ref{fig:ExperimentalSchematic})~\cite{thesisJohnson10}.
One of the particular challenges of an experiment using a long train (up to 10~\us) of very short pulses (10--20~\ns) is that the system is sensitive to both short- and long-time pulse distortions.
These effects included the intrinsic bandwidth of the AWG and the flux-pulse amplifier, the high-pass characteristics of the bias tee, a range of low-pass effects including the Minicircuits and eccosorb filters and filtering from the skin effect of the coaxial cabling, pulse bounces at impedance mismatches, as well as more intangible effects such as transient decays in step responses.
Subject to the system operating in a linear regime (e.g., the AWG operating in a comfortable amplitude range), this could be achieved by applying predistortions to the target fluxing sequence.

Figure~\ref{fig:FluxCals} illustrates the calibration process used in this experiment.
Rather than building a single, comprehensive model for all flux distortions, we took a divide-and-conquer approach, applying a series of corrections to compensate individual effects.
For processes outside the fridge, we calculated the required compensations by directly measuring the system step response using a fast oscilloscope (R\&S RTO1024, 10~Gs/s sampling rate and 2~$\GHz$ bandwidth).
We applied predistortion corrections sequentially, at each step correcting the longest-time behaviour and zooming in to shorter time scales once the longer-time response is successfully corrected.
Once measuring through the fridge, we optimised on the shape of the two-dimensional flux-pulse resonance, the so-called ``chevron''.
Again, we typically focussed initially on correcting the coarse features before zooming in to finer details.

The procedure we used to calculate the external corrections was:
\begin{itemize}
\item Sample a measured step response at a period $\tau$: $x[n] = x(n\tau)$.
\item Construct the system impulse response function according to: $h[n] = x[n] - x[n{-}1]$.
\item Construct the system transfer matrix $H$ from $h[n]$ ($H$ is a lower-triangular matrix with $h[j]$ in every position on the $j^{\rm th}$ lower diagonal).
\item Invert $H$ to find the transfer matrix of the so-called predistortion kernel and calculate the step response of the predistortion kernel as $H u[n]$, where $u[n]$ is the discrete Heaviside function.
This numerical matrix inversion step limits the length of the step response that can be treated in this way.
The sampling period $\tau$ is chosen to ensure the sampled step response covers the region of interest.
\item Fit the numerically inverted kernel step response using a simple functional form which can then be used to construct a high-resolution predistortion kernel (the impulse response calculated as above from a high-resolution step response).
The down-sampling of the step response reduces the fit function dependence on high-frequency effects.
For each step, we varied the sampling period to check that the fit parameters were relatively robust to details of the sampling.
\end{itemize}•

Figure~\ref{fig:FluxCals}(a) shows the step response from the AWG measured after the home-built flux-pulse amplifier (see Fig.~\ref{fig:ExperimentalSchematic}), with a zoom into the top of the step in (b).
In this case, the longest-time response was actually an effectively linear ramp over the long step response.
Here, we used a slightly modified procedure to the one above, fitting a linear function directly to the measured step response.
Using Laplace transformations, it is possible to show that a step response with a linear ramp, $(1+\alpha t) \, u(t)$, can be corrected using a predistortion kernel with an exponentially decaying step response $\exp(-\alpha t) \, u(t)$.  After this linear correction, we then implemented a series of three corrections with ``exponential-approach'' predistortion step responses of the form $(1+\alpha\exp(-t/\tau))u(t)$ with $\tau$ values between 5~$\us$ and 500~\ns\ (various amplitudes), determined using the above procedure.
Figure~\ref{fig:FluxCals}(c) shows the corrected step function measured after applying the four initial corrections.
The small but distinct sawtooth structure in the otherwise flat step response is due to the vertical resolution of the AWG.

After correcting for distortions from the AWG and flux-pulse amplifier, we measured the step response after the bias tee, at the fridge input.
Figures~\ref{fig:FluxCals}(d, e) show the measured step response and sampled predistortional kernel step response calculated using the above procedure (with $\tau=50$~\ns).
The high-pass characteristics of a reactive bias tee's RF input na\"ively predict a kernel step response with a full initial step followed by a continually increasing linear voltage ramp.
From Fig.~\ref{fig:FluxCals}(e), however, it is clear that the kernel step response is not completely linear.
We instead fit the step response to a quadratic form and proceed as above.
The step response measured after compensating for the bias tee is shown in Fig.~\ref{fig:FluxCals}(f).

Inside the fridge, we calibrated the flux-pulse predistortions to optimize the shape of the flux chevron [Figs~\ref{fig:FluxCals}(g--i)], which probes the excitation-swapping exchange interaction between qubit $\QR$ and resonator $\RR$ as a function of flux-pulse amplitude and interaction time.
When the qubit is exactly on resonance, the swapping interactions are expected to be slowest and strongest.
As it moves off resonance, the oscillations speed up and reduce in amplitude.
Interestingly, despite the good performance of the bias-tee correction when measured outside the fridge, the chevron measured with the same corrections [Fig.~\ref{fig:FluxCals}(g)] showed a clear ramp in the start of the interaction signal (the lateral skew), consistent with an under-compensated bias tee.
We do not understand the cause of this discrepancy, but corrected it empirically by adjusting the linear coefficient of the bias-tee correction.
The chevron measured after optimising this correction (final linear coefficient corresponded to a time constant $\tau=9.7$~\us) showed the characteristic asymmetric signature of low-pass filtering from the skin effect [Fig.~\ref{fig:FluxCals}(h)].
This was corrected by applying a kernel numerically calculated from a step response of the form $(1-\erf(\alpha_{\rm 1 GHz}/21\sqrt{t+1}))\, u(t)$~\cite{Wigington57}, using $\alpha_{\rm 1GHz}= 1.7$~\dB.
Finally, we implemented another series of exponential-approach kernels with values of $\tau$ between 1500~$\ns$ and 30~\ns, to achieve the result in Fig.~\ref{fig:FluxCals}(i).

\subsection{Calibration of the photon meters}
\label{sec:PhotonMeterCals}

\begin{figure*}[t]
  \centering
  \includegraphics[width=\linewidth]{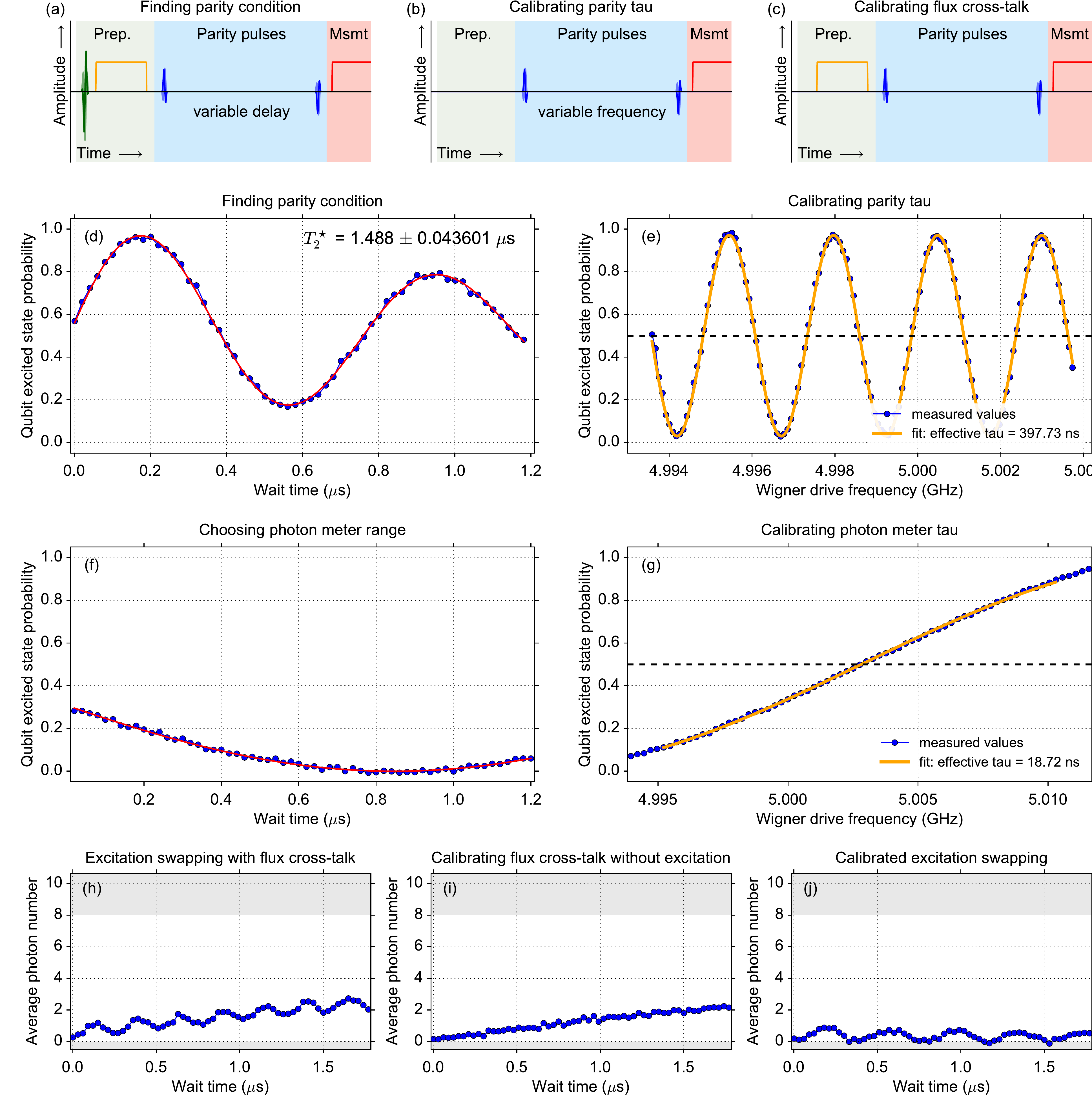}
  \caption{
  Calibration of the photon meter.
  (a--c) Measurement sequences used for calibrating the parity meter, specifically: (a) the dispersive shift of $\RR$ on $\QW$, (b) the effective delay time $\tau$ corresponding to a particular pulse separation, and (c) high-frequency flux cross-talk between flux pulses on $\QR$ and the flux offset of $\QW$.
  (d) Calibrating the parity condition,
  identified as the first crossing point of a Ramsey experiment with one photon in the resonator, giving a pulse separation of 383~\ns.
  (e) Calibrating the effective delay time $\tau$ for a particular pulse separation.  Using parity pulses separated by 383 $\ns$, we calibrated the effective separation $\tau$ to be 398~\ns, corresponding to a dispersive shift $2\chi/2\pi = -1.26$~\MHz.
  (f) Configuring an average photon number meter for a specific dynamic range of 0--8 photons.  Driving at the midpoint of the 0--8 photon frequency range, the Ramsey pulse separation is chosen to lie on the edge of the linear region.  For 0--8 photons, we chose to use a separation of 4 \ns.
  (g) Calibrating the photon meter effective $\tau$.  Repeating the measurement described in (e), the effective pulse delay for a 4 $\ns$ separation was $\sim 19$ \ns.  Comparing the oscillation period of the curves in (e) and (g) highlights the different sensitivity of the two photon meters.
  (h--j) Calibrating high-frequency flux cross-talk.  The flux cross-talk is calibrated by measuring the photon meter without loading excitations into the resonator and corrected by adjusting the phase of the second photon meter pulse.
}
  \label{fig:PhotonMeterCals}
\end{figure*}

Using a photon meter based on a Ramsey sequence's sensitivity to qubit frequency and $\QW$'s dispersive frequency dependence on resonator photon number allows detection of average photon number with controllable sensitivity and dynamic range.
Suppose the resonator is in the state $\psi\rr=\sum_j \alpha_j |j\rangle\rr$.
To implement the photon meter, we apply a Ramsey pair of $\pi/2$ pulses with pulse separation $\tau$ on $\QW$ at a frequency $\Omega\qw^d = \Omega\qw^0 - d 2\chi$, corresponding to the $d^{\rm th}$ photon peak.
Different photon-number frequency components accrue different phases during the variable delay between pulses, given by $\theta_j = (j{-}d) \, 2\chi \tau$.
By driving first around $\sigx$ and then around $\sigy$, the $d^{\rm th}$ photon term ends up on the equator of the Bloch sphere.
Measuring the excitation of $\QW$ then gives a measurement probability
\begin{align}
\nonumber
p^{\rm e}\qw = \sum_j \frac{|\alpha_j|^2}{2} (1+\sin\theta_j).
\end{align}
Provided $\tau$ is chosen such that $\theta_j$ is small for all photon components $j$ present in the photon state,
\begin{align}
\nonumber
p^{\rm e}\qw &= \frac{1}{2}\br{1+\sum_j (j-d) 2\chi\tau |\alpha_j|^2}, \\
\nonumber
&= \frac{1}{2}\br{1+ 2\chi\tau (\bar{n}-d)}.
\end{align}
Increasing $\tau$ therefore increases the sensitivity of measured probability to average photon number, but decreases the accessible range of photon numbers for which the linearity condition $\sin\theta_j \approx \theta_j$ holds.
An accurate calibration of the photon meter also requires an accurate calibration of the single-photon dispersive frequency shift $2\chi$ and $\QW$'s zero-photon frequency (which determines $\Omega\qw^d$).
Here, we describe a self-consistent calibration of our photon meters which does not rely on quantities derived from other measurements, such as spectroscopy, and relies primarily on knowing drive-pulse frequencies, probably the most accurate control parameter we have in the experiment.
At each stage, we first calibrate $\QW$'s zero-photon frequency using a standard Ramsey sequence.
With the performance of $\QW$ at the operating point (dephasing time $\Ttwostar \sim 1.5$~\us), we routinely achieved frequency accuracy better than 10~$\kHz$.

To calibrate the single-photon dispersive shift [sequence shown in Fig.~\ref{fig:PhotonMeterCals}(a)], a calibrated SWAP pulse on $\QR$ transfers an excitation into $\RR$, before the resonator photon number is probed via $\QW$.
The single-photon excitation in $\RR$ dispersively shifts the frequency of $\QW$ by $2\chi$.
Driving $\QW$ at the calibrated zero-photon frequency around $\sigx$ and then $\sigy$, the correct parity condition corresponds to the point where the curve crosses 0.5 excitation probability [Fig.~\ref{fig:PhotonMeterCals}(d): $383$~$\ns$ wait time].
This measurement is robust to both the relatively short resonator photon decay time $\Toneres\sim 3.5$ \us\ and the short dephasing time of $\QW$ at its operating point ($\Ttwostar\sim1.5$--1.8 $\us$ at $\sim -650$ $\MHz$ detuned from its top sweet spot), because these processes both reduce the visibility of the curve, but not the oscillation period, and therefore do not affect the value of the crossing point.
The zero-photon frequency calibration is the main limitation, because that calibration limits the accuracy with which the crossing point represents the correct delay time between $\pi/2$ pulses.

The wait time identified above specifies the time between the end of the first pulse and the beginning of the second required to realise a photon parity measurement, but this does not account for the finite pulse duration.
To calibrate the effective value of $\tau$, we fix the pulse separation and sweep the frequency of the $\QW$ drive generator this time without loading any photons into the resonator [Fig.~\ref{fig:PhotonMeterCals}(b)].
For a pulse separation of 383~$\ns$, the effective $\tau$ is $\sim398$~$\ns$ [Fig.~\ref{fig:PhotonMeterCals}(e)].
Note that the difference here is not quite the same as the drive pulse width used in the experiment ($4\sigma = 12$~\ns).
This value of $\tau$ is related to the dispersive shift of $\RR$ on $\QW$ in the usual way: $\tau=\pi/2\chi$, giving $2\chi/2\pi = -1.26$ \MHz.
Note that, when used directly as a parity meter, the read-out of $\QW$ was calibrated using a parity pulse pair either with the usual phase on the second pulse, or a phase shifted by $\pi$ radians.
This accounted for the reduced parity visibility from the short $\Ttwostar$ of $\QW$ at its operating point and helped to track any fluctuations in the correct parity extremes as a result of drift in qubit frequency and $\Ttwostar$.

Figures~2(a, b) show the pulse sequences for two different photon meters used in the experiment, one with the standard Ramsey sequence [calibrations in Figs~\ref{fig:PhotonMeterCals}(f, g)] and one an unbalanced ``echo''-like sequence with an off-centre refocussing pulse (calibrations not shown).
The mapping between average photon number and qubit excitation is approximately valid provided the phase advance/delay is less than 30 degrees, which corresponds to a qubit excitation of 0.25.
We select the appropriate Ramsey pulse separation by driving the qubit at the frequency corresponding to the mid-point of the desired range (here, the 4-photon position), calculated from the dispersive shift and the calibrated zero-photon frequency, and choosing the separation which gives the target excitation probability of 0.25 [Fig.~\ref{fig:PhotonMeterCals}(f)], here 4~\ns.
The effective $\tau$ was calibrated, as above, to be $\sim 19$~\ns.
Moving to the smaller $\tau$ necessary for a higher photon number dynamic range requires frequency refocussing.
Ultimately, the main limitation to the range achievable with such a photon meter is set by the bandwidth of the drive pulse.

We used a photon number meter calibrated using the above procedure to follow the excitation-swapping oscillations of a vacuum-Rabi exchange between $\QR$ and $\RR$, plotted as a function of the duration of the flux pulse on $\QR$ [Figure~\ref{fig:PhotonMeterCals}(h); sequence in Fig.~\ref{fig:PhotonMeterCals}(a)].
The drifting baseline results from pulsed flux cross-talk between $\QR$ and $\QW$.
To correct this, we repeated the same measurement without initially exciting $\QR$ in order to avoid exciting photons in $\RR$ [Fig.~\ref{fig:PhotonMeterCals}(c)].
This curve was compensated by adjusting the drive phase of the second Ramsey pulse in the photon meter (on $\QW$), leading to the compensated measurement in Fig.~\ref{fig:PhotonMeterCals}(j).
To maximise the sensitivity of the cross-talk calibration, during the calibration, $\QW$ can be driven at the zero-photon frequency, which then places the expected ``null'' measurement result on the equator of the Bloch sphere.
A modified version of this procedure can be carried out for all flux-pulse sequences of interest.
Note that cross-talk compensation was also necessary to ensure an accurate calibration of the parity condition in Fig.~\ref{fig:PhotonMeterCals}(d) above.

\subsection{Calibration of Wigner tomography}

\begin{figure*}[t]
  \centering
  \includegraphics[width=\linewidth]{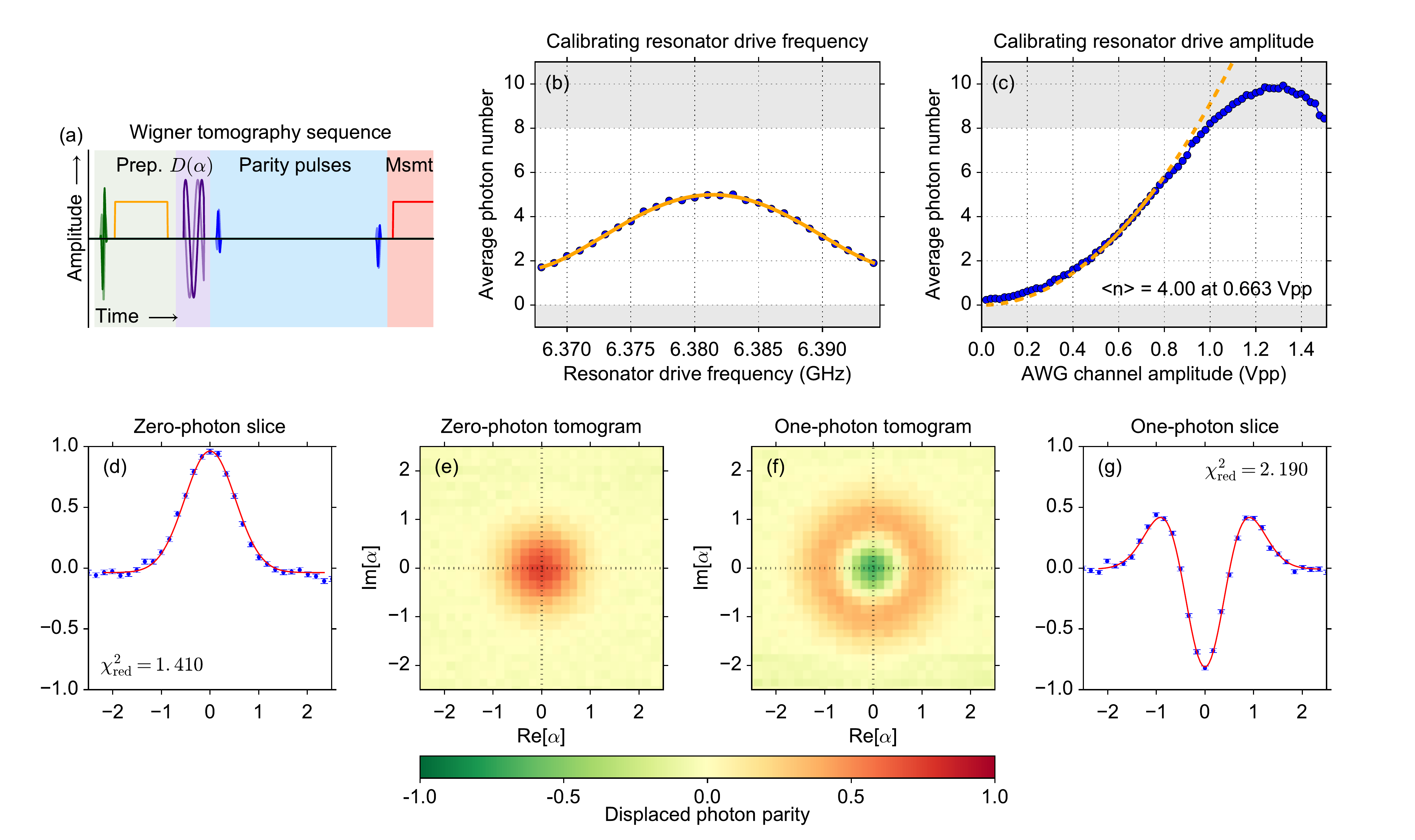}
  \caption{Calibration of Wigner tomography.
(a) Pulse sequence used to make the displaced photon parity measurement which provides a direct measurement of the Wigner function at a particular position in phase space.
(b) This plot shows the response of $\RR$ to the drive pulse as a function of drive frequency, as recorded by the $\QW$ photon meter, centred at 6.3814~\GHz, with a FWHM of $\sim 21$~\MHz, in reasonable agreement with the 18~$\MHz$ expected for a 50 $\ns$ square pulse.
(c) The pulse displacement amplitude is also calibrated using a low-dynamic-range photon meter with a linear range of 0--8 photons.  We fit the data in the centre of the linear range, where the photon meter mapping is most accurate, with a function of the form $\langle n \rangle = k_{\rm A} A^2$, finding $k_{\rm A} = 9.10$.
(e, f) Measured direct Wigner tomograms of zero-photon (e) and one-photon (f) states (one-photon state prepared using a calibrated SWAP pulse between $\QR$ and $\RR$).
(d, g) Direct Wigner tomogram slices of zero-photon (d) and one-photon (g) states measured using the full parity meter calibrations.
}
  \label{fig:WignerTomoCals}
\end{figure*}

We implement Wigner tomography using the direct method of Ref.~\onlinecite{Vlastakis13}.
After the algorithm part of the pulse sequence [represented in Fig.~\ref{fig:WignerTomoCals}(a) by a swap], a 50 $\ns$ square pulse applies a coherent displacement to the resonator photon state before the usual parity readout pulses.
The phase-sensitive resonator drive tone is created via single-sideband modulation in an IQ mixer.
We calibrate the drive frequency and amplitude using the already calibrated photon meter (Figs~\ref{fig:WignerTomoCals}(b, c), respectively).
The drive amplitude is calibrated in the middle of the linear range, where we expect the best performance.
Figure~\ref{fig:WignerTomoCals}(c) illustrates the breakdown of the linear mapping between average photon number and $\QW$ excitation probability both towards the edge of the linear regime and above the range, as the higher photon components wrap around in phase.
In the digital QRM simulation, for phase-sensitive Wigner tomograms (e.g., Figs~3 and 4), it was critical to maintain phase stability between the drives on $\QR$ and $\RR$ during the measurement.
To achieve this, the two microwave generators were synchronised using a 1~\GHz\ reference, with frequencies set as a multiple of the 5~\kHz\ experimental repetition rate.

Figure~\ref{fig:WignerTomoCals} shows one- and two-dimensional Wigner tomograms of a zero-photon (d, e) and one-photon (f, g) state (scaled in terms of photon parity).
The maximum visibilities in Figs~\ref{fig:WignerTomoCals}(f, g) do not reach the expected values, because these tomograms were measured without an accompanying full set of parity meter calibrations.
However, the radial symmetry observed in these tomograms demonstrates the correct behaviour of the coherent resonator drive.

The curves in Figs~\ref{fig:WignerTomoCals}(d, g) show fits to the data of a classical mixture of zero-photon and one-photon Wigner function cross-sections, with a free $x$-axis scaling parameter has been included in the fits.
These fits demonstrate that the measured tomograms agree well with theoretical expectations, subject to an $x$-axis scaling error of $\sim5\%$.
That is, the fits indicate that the amplitude calibrations result in a small systematic overestimate in displacement by 5\%.
This also agrees with two-dimensional double Gaussian fits of individual frames of the unconditional Wigner movie in Fig.~3(a) of the main text, which give an average Gaussian width $\bar{\sigma} = 0.526\pm0.003$, compared with the expected value of 0.5.

\subsection{Analog vs Digital Jaynes-Cummings Dynamics}

\begin{figure*}[t]
  \centering
  \includegraphics[width=\linewidth]{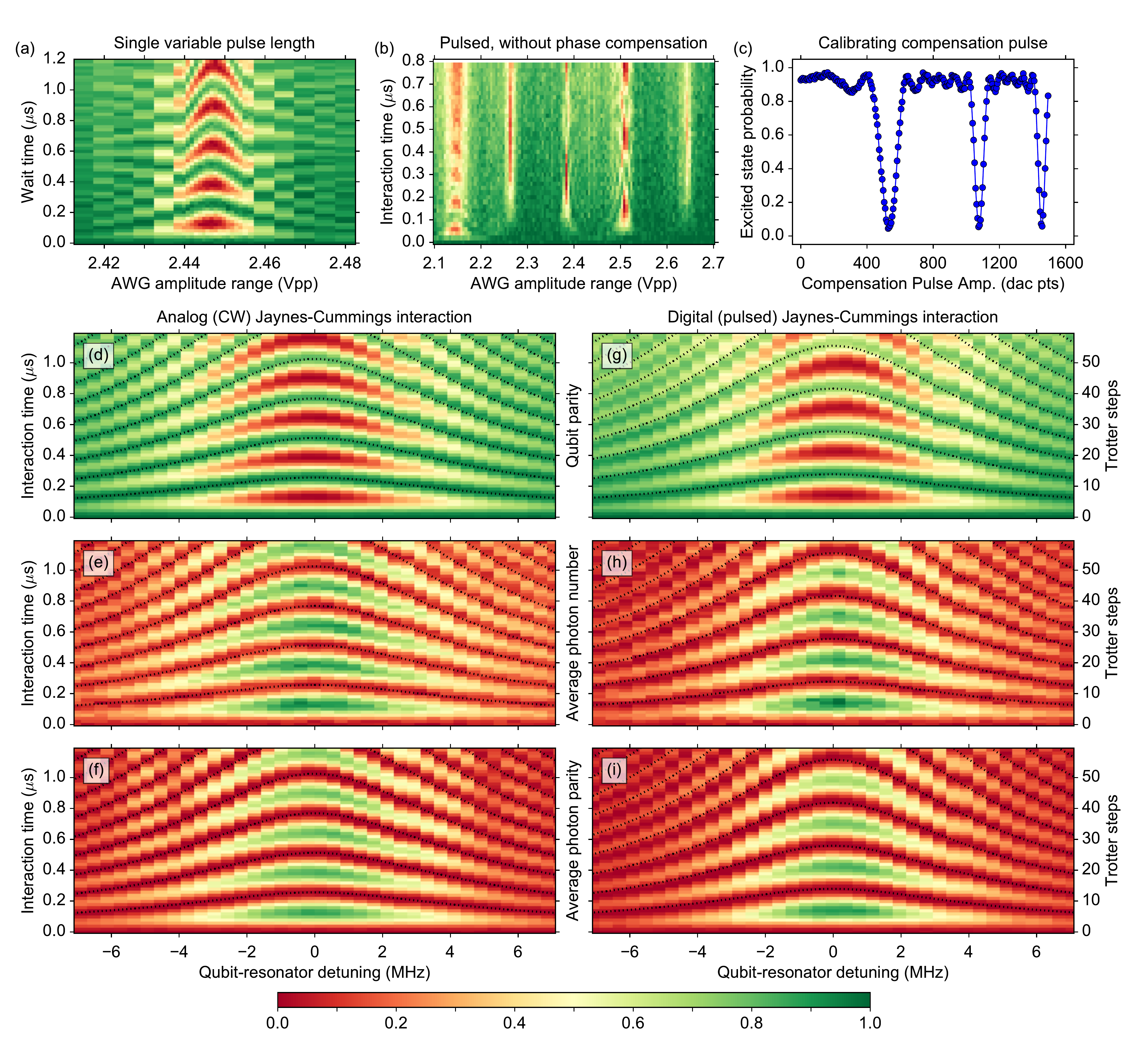}
  \caption{Comparison of analog and digital versions of a Jaynes-Cummings interaction.
  (a) Standard analog JC chevron showing the resonant excitation swapping between qubit and resonator after the qubit is initialised in the excited state as a function of flux-pulse amplitude ($x$ axis) and duration ($y$ axis) (qubit-resonator detuning and interaction time, respectively).  The $x$-axis location of the chevron ($\sim 2.445$ Vpp) therefore defines the qubit-resonator on-resonance condition.
(b) Digital JC chevron (measured under otherwise identical conditions) using a pulse duration of 20 $\ns$ showing a series of equally spaced resonances with different apparent interaction strengths.
(c) We scan amplitude of a 5 $\ns$ compensation flux pulse to identify the value which enforces that the digital chevron is centred around the natural resonance position.
(d-f) Standard analog and (g-i) digital JC chevrons measured by probing: (d, g) the excited state probability for $\QR$, (e, h) the average photon number in $\RR$ (linear range 0--2 photons), and (f, i) the photon parity of $\RR$.
}
  \label{fig:CWvsPulsedJC}
\end{figure*}

Simple modelling of the Trotterised version of the full Rabi model shows that high-quality simulations require both slow dynamics and short Trotter steps (i.e., fast flux pulsing).
Such an experiment is sensitive to both short-time and long-time effects in the flux-pulse shaping.
A simpler experiment which verifies the performance of this flux pulsing is to implement a digital simulation of the standard Jaynes-Cummings (JC) interaction underlying the standard excitation-swapping experiments demonstrated with single flux pulses~(Fig.~\ref{fig:FluxCals}).

In the standard continuous-wave (single-pulse) version of a JC excitation-swapping interaction, resonance between the qubit and resonator frequencies gives rise to maximum visibility oscillations of the excitation moving between the two components.
When detuned, the different phases accrued by the qubit and resonator during the interaction decrease the oscillation visibility, while increasing the oscillation frequency.
This gives rise to the characteristic shape of the flux chevron.
Significant care is required, however, to accurately reproduce the (analog) JC interaction with a digital pulse train.

Figures~\ref{fig:CWvsPulsedJC}(a, b) show analog and digital versions of the JC interaction (viewed through the qubit excitation) under otherwise identical conditions.
The digital chevron shows a series of resonances which do not appear in analog measurements (not shown), and there is also no chevron visible at the natural resonance condition around 2.45 Vpp.

The new features relate to the extra ``interaction off'' times in the digital version.
The regular spacing between neighbouring satellite resonances is around 50 \MHz\ (after converting AWG amplitude to qubit frequency), which is the inverse pulse duration.
During the interaction time, the qubit-resonator relative phase evolves as expected.
However, in the ``off'' time between interaction pulses, the qubit accrues phase at a different rate, and will hence not have the required phase at the beginning of the next pulse for the interaction to pick up where it left off at the end of the previous pulse.
Therefore, the necessary condition for observing a chevron feature at exactly the position of the natural resonance is that the qubit phase accrued (relative to the resonator) during the ``off'' time should be a multiple of $2\pi$.
The observation of multiple satellite resonances is a form of digital aliasing, where the interaction will build up constructively from pulse to pulse provided the relative phase accrued between qubit and resonator during the ``on'' time of the pulse again differs only by an integer multiple of $2\pi$.
However, this is an aliasing of the dynamics itself, not just an aliasing of the measurement, which could also occur in natural continuous-wave (CW) chevrons and would never lead to the observation of extra satellite peaks.

This pulsed interaction can also be viewed as a Trotterised simulation of the CW interaction.
While successive interaction pulses obviously commute with each other, they do not necessarily commute with the ``off'' pulses.
The condition on qubit-resonator phase during the ``off'' pulse can be understood as the condition where the Trotter error vanishes, because the Hamiltonian term resulting from the qubit detuning coincides with the identity.
The satellites arise because the phase contribution from the qubit detuning in the ``on'' pulse is identical if the frequency change matches a multiple of $2\pi$ phase.

To compensate for the phase error accrued in the qubit during the ``off'' pulses, we apply a 5 $\ns$ compensation flux pulse between interaction pulses.
Using the flux-pulse amplitude which corresponds to the centre of the CW chevron, the amplitude of the compensation pulse was swept to identify the correct compensation point.
In this way, very good agreement was achieved between the digital JC dynamics and the traditional analog version [Figs~\ref{fig:CWvsPulsedJC}(d--i)].
The main differences are a slightly reduced visibility because of the increased experiment time, and a slighly lower effective coupling frequency ($g/2\pi\sim1.8$ \MHz, instead of $\sim1.95$ \MHz).
The latter most likely arises from residual short-time pulse imperfections which do not contribute significantly to the long interactions in the analog form.

\subsection{Trotter simulation with excited and ground initial states}

\begin{figure*}[t]
  \centering
  \includegraphics[width=\linewidth]{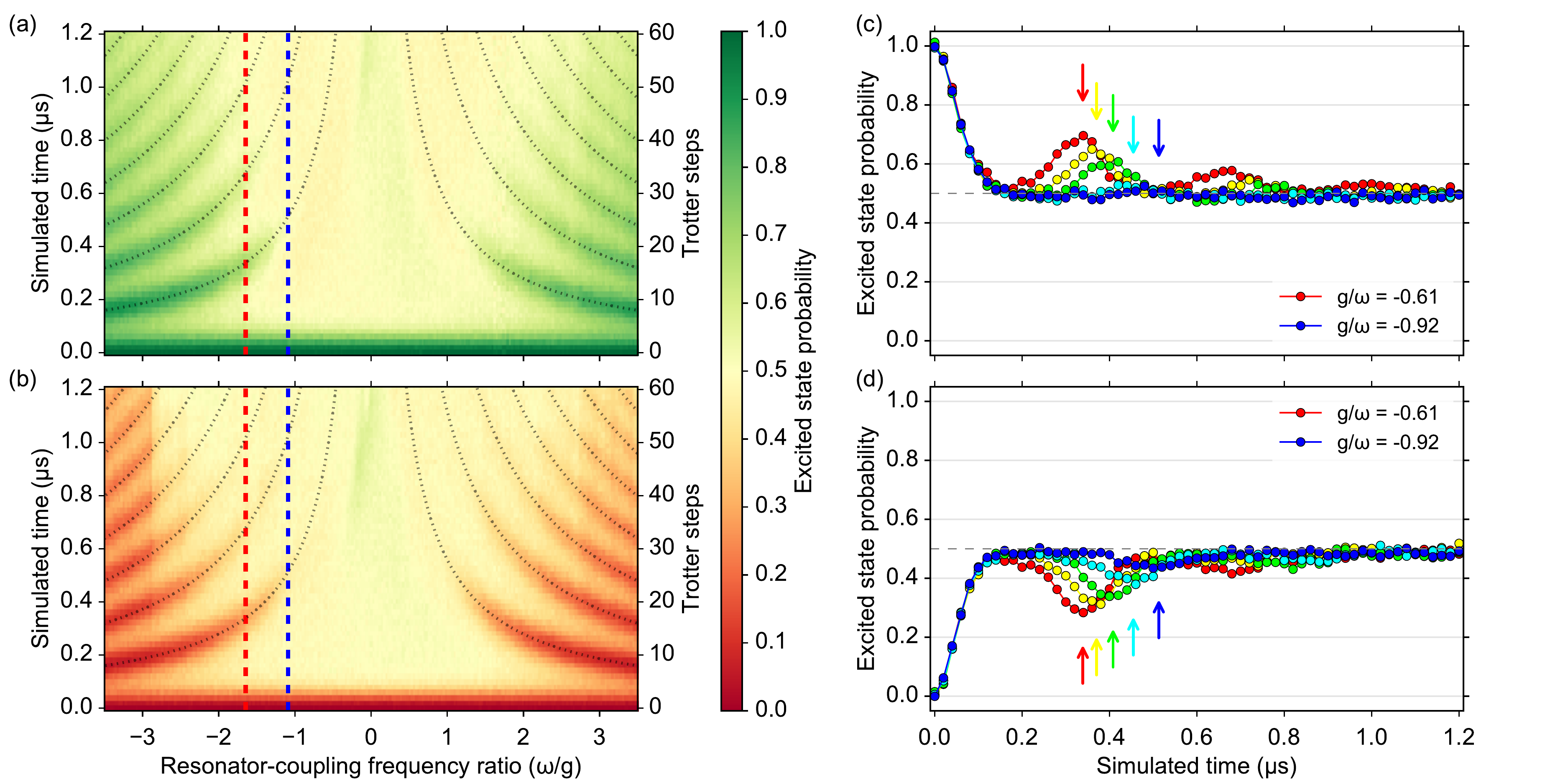}
  \caption{
  Comparison of the simulation between initialising in the excited (left) versus the ground state (right).
  (a, b) These plots directly verify the symmetrical behaviour of the simulated Rabi model.
  (c,d) Line slices are plotted at evenly spaced frequencies between the red and blue dashed lines in (a, b).  Arrows in (c, d) show the expected time for the first revival.
  }
  \label{fig:ExcitedVsGround}
\end{figure*}

  In the degenerate-qubit case, when understood in terms of the cavity trajectories in phase space, it is clear that the structure of the dynamics of the full Rabi model with USC should not depend on whether the qubit starts in the ground or excited state.
  This contrasts with the JC interaction, where the $|g,0\rangle$ state is decoupled from the rest of the system and the system will only undergo nontrivial dynamics if an initial excitation is loaded in the system.
  Indeed, in a natural USC system, if it were possible to turn the coupling on and off rapidly, it would be extremely interesting to watch an uncoupled-system ground state evolve into a state with excitations in the qubit and cavity.
  In this digital simulation, however, this is less satisfying, since the protocol in any case involves regularly injecting excitation into the system in the form of qubit flipping pulses.
  Most of the results reported here therefore take the more conservative position of initialising the system with an excitation, with the motivation that observing a difference between the simulated dynamics and what would be expected in a weak-coupling scenario could then only result from the simulated counter-rotating terms.
  Although there were some stability issues during the measurement with ground-state initialisation, there is nevertheless extremely good agreement between the two cases, for example with the timing of the revivals in both cases agreeing with the theoretical predictions.
For this particular measurement of ground-state initialisation, qubit revivals are observed even out to $\gomratio\equiv\gR/\omqR \sim 1$.

\subsection{Trotterisation performance vs Trotter order}

\begin{figure*}[t]
  \centering
  \includegraphics[width=\linewidth]{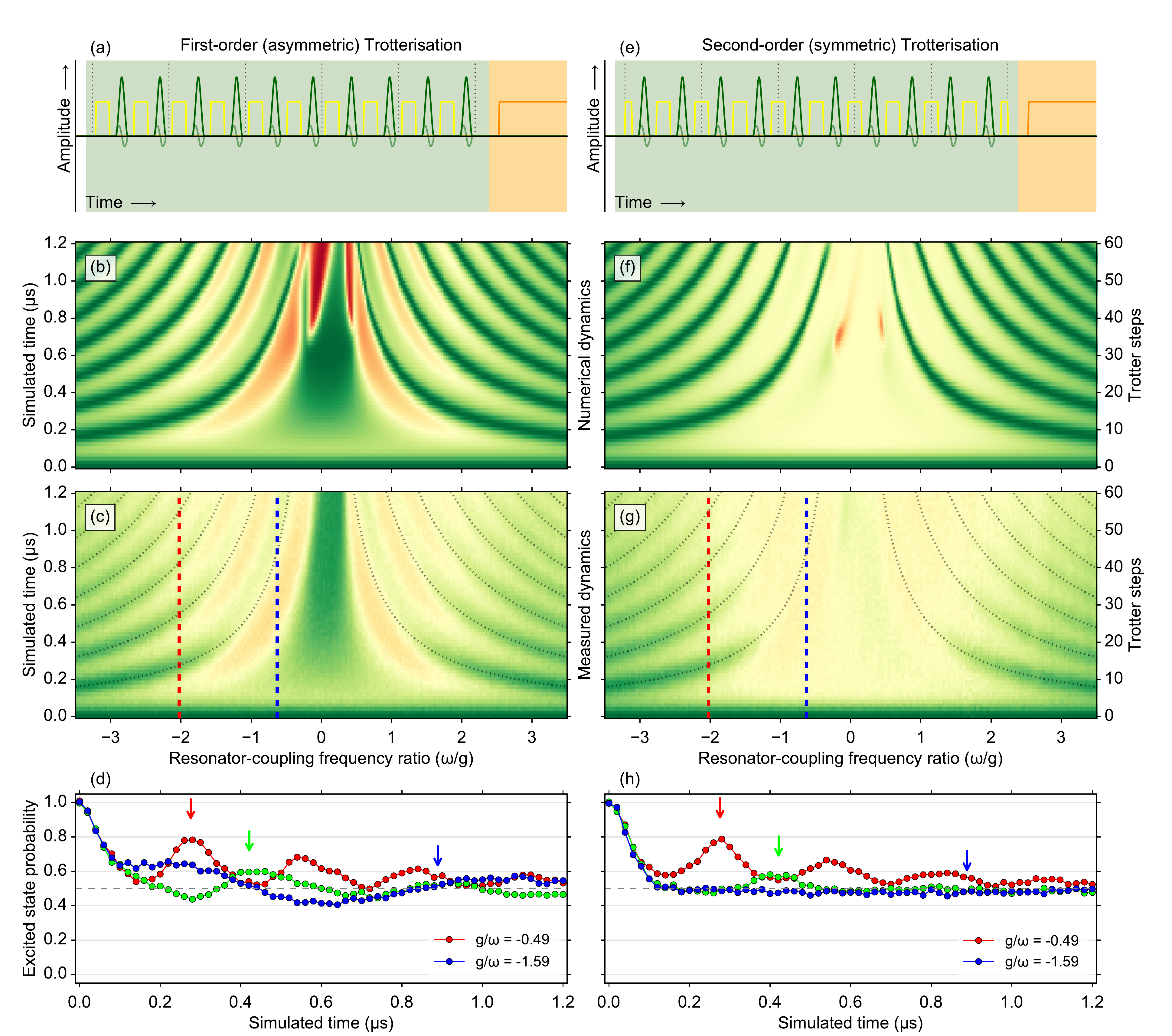}
  \caption{
  Comparison of simulation performance for asymmetric, first-order (a--d) and symmetric, second-order (e--h) Trotterisation.
  (a, e) Pulse sequences for the first-order (a) and second-order (e) Trotterisation.
  (b, f) Numerical simulations of the Trotterised Rabi model for the ideal case with no decay.  Note that the sharp features in the centre of the plots (deep in the ultrastrong coupling regime) are not artifacts of the numerics, but Trotter error related to the 20 ns step size (these features disappear for 10 ns pulses).
  (c, g) Experimental quantum simulations for first-order (c) and second-order (g) Trotterisation, showing very good agreement with the numerical results in (b, f).
  (d, h) Vertical line slices are plotted for evenly spaced resonator frequencies between the red and blue dashed lines in plots (c) and (g).  
  }
  \label{fig:SymVsAsym}
\end{figure*}

As discussed already, initial modelling of a Trotterised Rabi simulation showed that unusually low qubit-resonator coupling between \QR\ and \RR\ was required to be able to achieve reasonable simulation fidelities given the hard bandwidth limitations of flux-based fast frequency tuning.
This, however, required longer experimental times for the simulations, which in turn placed significant constraints on qubit and resonator coherence.
Indeed, the shorter-than-anticipated resonator coherence time proved to be the biggest limitation.
As a result, it was critical to use all available measures to minimize the Trotter error in our simulations, given the limits on the shortest achievable Trotter step sizes.

The accuracy of the Trotter approximation is set by the amount of non-commutativity between different components in the step~\cite{Lloyd96SOM}.
While first-order Trotterisations [$\exp(A+B) \approx \exp (A) \exp (B)$] lead to Trotter errors that scale with single commutators (quadratically with simulation time), higher-order Trotterisations can be used to eliminate lower orders of Trotter error.
For example, the symmetry of a second-order Trotterisation [$\exp(A+B) \approx \exp (A/2) \exp (B) \exp (A/2)$] ensures that first-order error terms (related to single commutators) cancel, pushing the largest Trotter error terms out to third order in simulation time.
For two-part Hamiltonians, however, second-order Trotterisation in practice only involves modifying the pulses in the first and last Trotter steps.
All the results in the main text were obtained using a second-order Trotterisation.
The plots in Fig.~\ref{fig:SymVsAsym} illustrate that this was absolutely critical in order to extend the simulations deep into the ultrastrong coupling regime.
  The first-order and second-order Trotterisation agree reasonably well at $\gomratio<0.5$, but behave fundamental differently at the higher values.
  The first-order simulation starts to show qualitatively different behaviour for relative coupling strengths $\gomratio\gtrsim0.5$.
  In particular, only in the second-order case are the characteristic plateaus and revivals of the USC regime observable.

\subsection{Trotterisation performance vs Trotter step size}

\begin{figure*}[t]
  \centering
  \includegraphics[width=\linewidth]{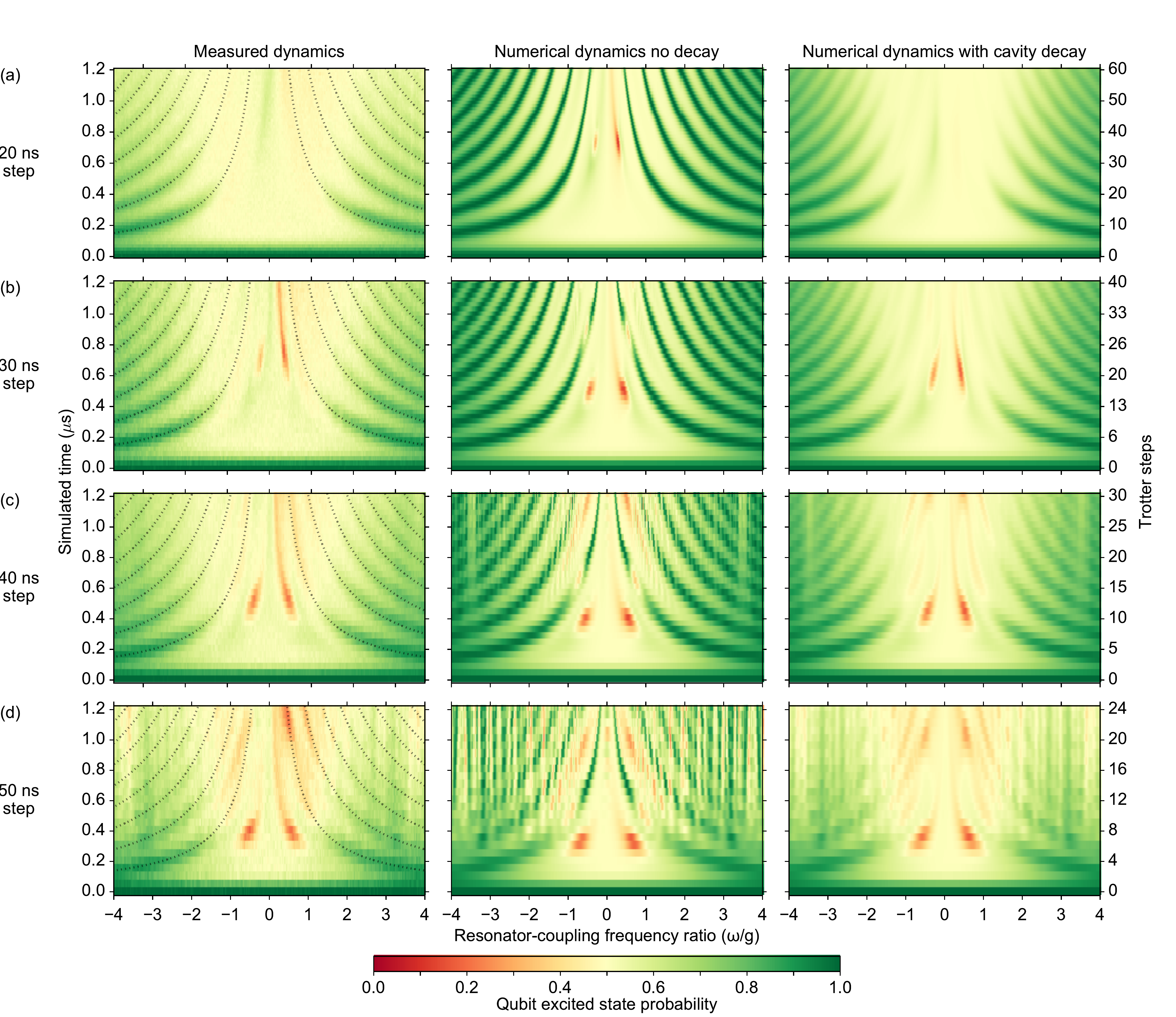}
  \caption{
  Comparison of simulation performance for various Trotter step sizes, showing measurements (left), numerical simulations with no decay (middle) and numerical simulations with the measured $\Toneres=3.5$ \us: (a) 20 $\ns$ steps (60 Trotter steps), (b) 30 $\ns$ steps (40 Trotter steps), (c) 40 $\ns$ steps (30 Trotter steps) and (d) 50 $\ns$ steps (24 Trotter steps).
  }
  \label{fig:TrotterStepSize}
\end{figure*}

  As illustrated in Fig.~\ref{fig:SymVsAsym}, the effects of Trotter error are most visible in the high $\gomratio$ regimes, which is reasonable, considering that for low $\gomratio$, the Rabi model is well approximated by the JC model where the excitation-nonconserving terms (non-commuting with the excitation-conserving terms) do not play a significant role.
  This was also visible when studying the performance of the simulation as a function of the Trotter step size.
  
  Measurements and numerical simulations show significant reduction in Trotter error as the number of Trotter steps over 1.2 $\us$ increased from 24 to 60.
  The Trotter error shows up in two ways, namely the central features departing from the expected plateaus, and a tendency for the dynamical landscape to ``break apart'', even out into the lower coupling regimes.
  In the measured results and the simulation with decay, the fine details do not appear as strongly, but the effect appears to wash out the oscillation dynamics more rapidly.
  Only at the smallest step size are these effects absent from the measured results, and in the ideal simulations (without decoherence) there are even then central features which only disappear at a still smaller 10 $\ns$ step size.
  The measured results agree very closely with the numerical Trotter dynamics which include only the effect of photon decay, again highlighting that the primary limiting factor in our experiments was $\Toneres$.
  It is clear from these results that moving towards the smallest possible Trotter steps will be a key challenge for reaching quantum supremacy in complex quantum simulations.

\subsection{Qubit entropy dynamics}

\begin{figure*}[t]
  \centering
  \includegraphics[width=0.6\linewidth]{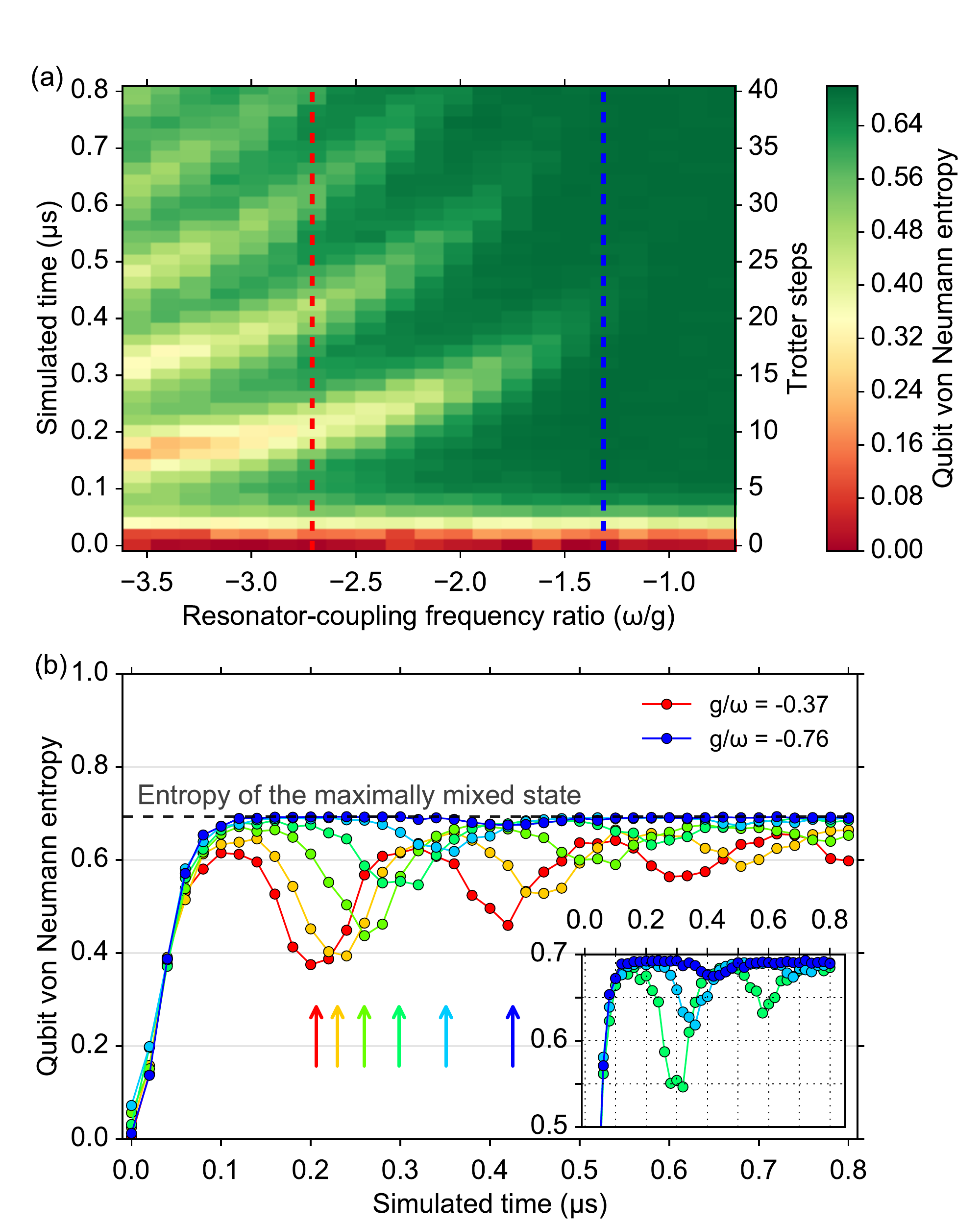}
  \caption{
  Tomography of the reduced state of qubit $\QR$ as a function of simulation time and relative resonator-coupling frequency.    (a) Image plot showing the dynamics of qubit quantum von Neumann entropy over different ultrastrong-coupling regimes.
  (b) Line slices are plotted at evenly spaced frequencies between the blue and red dashed lines.
  Inset: Zoom showing revivals.
  }
  \label{fig:TomoChevron}
\end{figure*}

  In the Rabi model, as the resonator states separate, the qubit-resonator entanglement causes the reduced qubit state to collapse towards the maximally mixed state.
  A revival occurs in the qubit purity only if the underlying entanglement is still present when the resonator states re-coalesce at the origin in phase space.
  While many possible uninteresting effects may cause an initial collapse in qubit purity, a revival in purity is a signature of entanglement with another system, in this case the resonator.
  After each Trotter step, a tomographically complete set of measurements on $\QR$ was used to reconstruct its reduced state using maximum-likelihood tomography.
  We use the von Neumann entropy to characterise the purity of the reduced qubit state and observe revivals in qubit purity out to $\gomratio > 0.8$ [Fig.~\ref{fig:TomoChevron}(a)], consistent with the observed revivals in qubit parity.
  While the observed revivals shown in the slices [Fig.~\ref{fig:TomoChevron}(b)] appear smaller than the qubit parity revivals, in fact this is deceiving, resulting from the fact that purity (as with other entropy measures) is a quadratic function of the qubit population difference.
  The inset shows that the background noise of this signal is small and that the revivals are quite distinct.
  Moreover, plotting an appropriate square root of the entropy (not shown) shows that the revivals are consistent with the qubit parity case.

\end{document}